\documentclass[aps,prl,twocolumn,showpacs,superscriptaddress,groupedaddress]{revtex4}
\usepackage{graphicx}  
\usepackage{dcolumn}   
\usepackage{bm}        
\usepackage{amssymb}   
\usepackage{slashed}
\usepackage{mathtools}
\usepackage{multirow}
\usepackage{nameref}
\usepackage{varioref}
\usepackage{hyperref}
\usepackage{cleveref}

\begin{document}

\title{Reconstructing the electro-weak theory from a pure gauge theory}
\author{L. Yang}
\thanks{yanglu@xaut.edu.cn}
\affiliation{Department of Applied Physics, Xi'an University of Technology, Xi'an 710054, China}


\date{\today}

\begin{abstract}
$U(4)$ local transformations on the four Weyl spinors forming the isospin doublet of Dirac fermions are assumed as symmetries of the standard model. With the Lorentz transformations considered simultaneously, the symmetry group is enlarged in order to form a closed Lie algebra. In this framework, the chirality mixing gauge components with certain constraints collectively are identified as the Higgs field in the standard model. The scalar-appearance of the gauge-natured Higgs field and its varying coupling constants with the fermions, i.e. the mass-ratio parameters, are also given natural explanation. Additionally, a background of a constant right-handed gauge component is postulated to obtain the symmetry-breaking Higgs potential, which also leads to a possible explanation to the parity violation in weak interactions. Further, the new framework provides an alternative perspective to understand the hyper symmetry $U(1)_{Y}$, as a combination of a simultaneous phase rotation on both the left and right handed fermions and a rotation in the third direction of $SU(2)_{R}$ on the right-handed fermions. At last, a topological term involving the Higgs field is found to exist. 
\end{abstract}

\maketitle

\section{Introduction}\label{sec:intro}
The standard model of particle physics is by far the most successful theory. The development of the theory has gone through a series of conceptual and theoretical breakthroughs, such as the symmetry-breaking concept introduced to particle physics by Nambu, the Higgs mechanism innovated by Higgs, Englert et.al, the proof of the renormalisability of Yang-Mills theories by 't Hooft and Veltman, and the electro-weak unification model invented by Glashow, Weinberg and Salam, only to mention the most influential and relevant. Physicists' confidence in the standard model was consolidated in 2012 when the predicted Higgs particle was eventually discovered. The extensive experimental research before and after the discovery also accumulated even more faith in the standard model. Despite the great success, there are still some concerns about the CP violation, the value of muon $g-2$ and the Higgs-mass problem. These concerns of course do not tarnish the glory of the standard model at all. But the desire for removing every single piece of doubt encourages physicists not to cease in searching for new physics. 

In the standard model the most confusing part is perhaps about the Higgs sector. It originates from a concept borrowed from the famous BCS theory of low-temperature super-conductance first introduced by Nambu. In BCS theory, the symmetry-breaking potential comes from the lattice-mediated interaction between the electrons that produce an effect of weak attraction between the electrons. Despite the complexity, the theory provides a concrete and natural explanation for the mechanism. In the standard model, on the opposite side, the dynamics of the Higgs field comes in a heuristic and artificial way. Another aspect concerns the understanding of the Yukawa terms involving the Higgs field: does the Higgs field represent a fifth type of interaction, besides electromagnetic, weak, strong and gravitational interactions? Given the strong similarity between the Higgs field and a non-abelian gauge field component, it is tempting to speculate that the Higgs field might indeed be some gauge component. This speculation finds support in a number of remarkable researches. The first support comes from Witten's accidental discovery \cite{Witten} that the $SU(2)$ gauge theory in 4-spacetime reduces to an abelian Higgs model, upon imposing spherical symmetry. The resulted abelian Higgs model has the correct covariant derivative and symmetry-breaking quartic potential, except that the underlying spacetime effectively becomes a $1+1$-d curved spacetime. Inspired by this work, Manton made the first attempt to derive the Weinberg-Salam model from a pure gauge theory \cite{M}. The second support is the general analysis completed by Forgac and Manton\cite{MF} showing that this reduction of a big gauge group to a certain subgroup as the symmetry group of the reduced theory with a Higgs like scalar field, upon imposing spherical spacetime symmetry, is generic. These two discoveries strongly hint that the Higgs field and gauge fields must be deeply connected to each other. The third support comes from Julia and Zee's observation\cite{JZ} that the temporal component of a time-independent gauge field can be viewed as a Higgs field with natural covariant derivatives. This observation did not receive sufficient attention in the literature, but turns out to point to a bright direction in searching for the true identity of the Higgs field. These pioneering researches lend strong confidence to the idea that the Higgs field has a gauge origin and encourage us to continue the journey to explore in this direction. 

If our attention is directed to the fermion-gauge coupling terms instead of getting the Higgs sector in the Lagrangian right as the primary goal, Julia and Zee's observation guides us to look at the temporal component of the gauge field. This inspires us to put the Yukawa term involving the actual Higgs field in the same form as the temporal gauge field. At the end this leads us to relate the Higgs field to the chirality mixing gauge components. They are matched comfortably! However, there are still two serious obstacles when we attempt to identify the Higgs field as the chirality mixing gauge components. First, a non-abelian gauge theory can have only one coupling constant. But the Higgs field in the standard model couples to the fermions with different coupling strength (fermion mass ratios). This conflict will be referred to as the mass-ratio problem in the later text. The resolution is found to be the Lorentz transformations with the hypothesis that the usual Lagrangian has been Lorentz transformed in a special way. In short, a boost transformation rescales the coupling strength and a rotation redistributes the weight in mass ratios of the two fermions in a generation. Further, for different generations, the boost and rotation transformations are different but compatible. Thus the fermion mass ratios can be understood as Lorentz charges in the same sense as the electric charge to the $U(1)$ electric-magnetic transformation. The second obstacle is that a gauge field is a vector field while the Higgs field behaves as a scalar field. The resolution of this conflict relies on the remarkable interweaving between the Dirac gamma matrices, the chirality mixing generators and the Lorentz transformations. Upon a Lorentz transformation, the product of gauge components and the Dirac gamma matrices transform only in the internal space. When the chirality mixing gauge components are excited with a certain constraint, the product is collectively invariant under any Lorentz transformation. This makes the Higgs field appear as a scalar. 

With the chirality mixing gauge components identified as the Higgs field, it is possible to reconstruct the abelian Higgs model and non-abelian Higgs model starting from pure gauge theories. For this the chirality mixing gauge field components are assumed to take special forms, i.e. subjected to certain constraints, in the same sense as imposing spacetime symmetries as in \cite{Witten} and \cite {MF}. In addition, a constant background of a right-handed gauge component is postulated to generate a non-zero vacuum expectation value of the Higgs field. The reconstruction leads only to a minor change in the abelian case. But in the non-abelian case, this offers a number of surprising new perspectives to look at the electroweak sector in the standard model. First, the hyper $U(1)_{Y}$ is reinterpreted as generated by a common phase transition to both the left and right handed fermions, followed by the same phase transition only to the right handed fermions. This interpretation implies that the Weinberg angle $\theta_{w}$ takes the value $\arctan \frac{\sqrt{3}}{3}$ can be deduced as a necessary consistent condition. Second, the right-handed constant background leads also to massive bosons associated to the right-handed gauge fields. Their masses are estimated to be at least 1.9 times that of the left-handed gauge bosons. This might be able to explain the parity violation observed in weak interactions. The third observation is that when the Higgs field is identified as a gauge component, a topological term involving the Higgs field can be constructed. The reconstruction in the non-abelian case takes only the weak coupling constant and the fermions mass ratios as input. We have not been able to determine the fermion mass ratios from a geometric or algebraic way, albeit such possibilities are not impossible.

This paper is organised as follows: after the introduction, in the second section the algebras of the symmetry group of the Lagrangian to be considered are specified and the gauge field components are introduced. Then the special properties of the chirality mixing gauge components when coupling to the fermions are discussed, to show why they may be identified as the Higgs field. In the third section, the invariance of the chirality mixing gauge components in certain special situation is shown, to explain why the Higgs field appears as a scalar. The mass-ratio problem is then addressed and the resolution is explained. In the meantime, a consistent theory is defined with each generation of fermions experiencing different Lorentz transformations. In the fourth section, special gauge configurations are postulated and the abelian and the standard model Higgs theories are reconstructed and discussed. Then the paper is ended with discussion on the remaining problems and further possible development. 

Throughout the greek letters $\mu, \nu, \alpha, \beta$ and the latin letters $j, k, l$ are employed to index both the spacetime directions as well as the 4 rank-2 unitary matrices $I_{2}:=\sigma^{0}$ and the Pauli matrices $\sigma^{1}, \sigma^{2}, \sigma^{3}$. The greek letters run in the range $0,1, 2, 3$ and the latin letter run in the range $1, 2, 3$. In the context of spacetime $0$ denotes the time direction. Overall summation is implied upon double occurrence of dummy indices unless otherwise stated. Please also be aware that although all transformations are denoted by $U$, they can be either unitary or non-unitary.

\section{The Higgs Field as Gauge Components}
The modern approach to construct an interacting theory follows Weyl, Yang-Mills and Utiyama \cite{WFYU}. One starts with a non-interacting theory and identify the global symmetry transformations. Then these symmetry transformations are made local, i.e. to be allowed to depend on spacetime locations. Once the symmetry transformations depend on spacetime location gauge potentials are necessary to present in order to keep the theory invariant under the local transformations. The gauge potentials couple to the fermions and mediate various interactions between the fermions. Nowadays the concept that interaction comes from local symmetry transformations, or are geometrically identified as connections and curvatures, has deeply rooted in physicists' mind and commonly known as gauge principle. Utiyama in 1956 set up the framework to make this concept mathematically clear. He was also the first to get gravity treated in this framework. Today, the standard model of particle physics is the most successful gauge theory. But the Higgs field in the theory when viewed as an interaction does not fall in the gauge framework. In the following let us make an attempt to put the Higgs field and the gauge field in a unified gauge framework. 

\subsection{The Lagrangian and the symmetry transformations} \label{sec:symmetry}
Let us start with the Lagrangian for a pair of non-interacting massless Dirac fermions (this can be easily extended to include 6 pairs of Dirac fermions, the three pairs of leptons and another three pairs of quarks as in the electro-weak interactions):
\begin{equation}  
L= \bar{\psi}_{1} i \slashed \partial \psi_{1}+  \bar{\psi}_{2} i \slashed \partial \psi_{2}
\end{equation}
where $\bar{\psi}_{s}=\psi_{s}^{\dagger}\gamma^{0}$ ($s=1,2$), and the Feynman symbol $\slashed{\partial}$ represents the contraction of the partial derivative and the Dirac matrices, i.e. $\slashed{\partial}= \partial_\mu \gamma^{\mu}$. A massless Dirac fermion is equivalent to two Weyl fermions, one left handed and one right handed. So there are four Weyl fermions in this theory, $\psi_{1L}, \psi_{1R}, \psi_{2L}, \psi_{2R}$. For convenience let us rewrite the Lagrangian as:
\begin{equation} 
L=\bar{\Psi} i \slashed \partial \Psi
\end{equation}
where
\begin{align}
\Psi&=\begin{pmatrix}
\psi_{1R}\\
\psi_{2R}\\
\psi_{1L}\\
\psi_{2L}
\end{pmatrix}
\end{align}
and  where now $\bar{\Psi}=\Psi^{\dagger}\Gamma^{0}$, $\slashed \partial = \Gamma^{\mu} \partial_\mu$ with
 \begin{equation} \label{eq: Gamma}
 \begin{split}
 \Gamma^{0}=&\sigma^{1}\otimes I_{2}\otimes I_{2}\\
 \Gamma^{k}=&i \sigma^{2}\otimes I_{2}\otimes \sigma^{k}
 \end{split}
 \end{equation}
Further $\Gamma^{0}\Gamma^{0}=I_{8}$ and $\Gamma^{0}\Gamma^{k}=-\sigma^{3}\otimes I_{2}\otimes \sigma^{k}.$ Let's assume that this Lagrangian admits the symmetry transformations of $U(4)$ among the four massless Weyl fermions:
\begin{equation}
\Psi \rightarrow U\Psi.
\end{equation}
The generators of $U(4)$ are normally given by $\sigma^{\mu}\otimes\sigma^{\nu}\otimes I_{2} $ with $ \mu, \nu \in \{0,1,2,3\}.$ Let us linearly combine them into three sets:
\begin{equation} \label{eq: gen_U(4)}
\begin{split}
g_{4R}&:=\{ \frac{1}{2}(I_{2}+\sigma^{3})\otimes \sigma^{s}\otimes I_{2} \} :=\{ \sigma^{R}\otimes \sigma^{s}\otimes I_{2} \}\\
g_{4L}&:=\{\frac{1}{2}(I_{2}-\sigma^{3})\otimes \sigma^{s}\otimes I_{2} \} :=\{ \sigma^{L}\otimes \sigma^{s}\otimes I_{2} \}\\
g_{4M}&:=\{ \sigma^{1,2}\otimes \sigma^{s}\otimes I_{2} \} ,s=0, 1,2,3
\end{split}
\end{equation}
The transformations generated by $g_{4R}$ ($g_{4L}$) have obvious physical meaning as $U(2)$ transformations acting on the two right (left) handed fermions, i.e. they keep the handedness unchanged. They will be referred to as chirality keeping transformations. Those transformations generated by $g_{4M}$ mix the left- and right-handed fermions, they will be referred to as chirality mixing transformations. The chirality keeping transformations commute with the product $\Gamma^{0}\Gamma^{\mu}$, while the chirality mixing ones do not. The non-commutativity causes problem when viewing the chirality mixing transformations as symmetries of the Lagrangian, the solution to this problem will be given shortly. 

The fact that the group generators and the Gamma matrices act on the same spinors consecutively suggests that these generators and the Gamma matrices, as well as the generators of the Lorentz group should all be treated in the same context of a Clifford algebra, which would give the product of any two of these matrices a natural geometric meaning. In the meantime, the $U(4)$ transformations and the Lorentz transformations when assumed to belong to the same group, their generators should also belong to the same Lie algebra. On a geometric algebra an extra Lie-algebra structure can be naturally imposed. So there is no conflict for both viewpoints. When the target set is viewed as the Clifford algebra, its generating basis contains of the four Dirac matrices. Following the generating rule, the full algebra can be at least partially generated. When the target set is taken to be  a Lie algebra, we can only start with the generators of the Lorentz group and those of $U(4)$ to generate the smallest Lie sub-algebra. The best hope is that these two sets are isomorphic to each other. But there is no guarantee.  

The four Dirac Gamma matrices $\Gamma^{\mu}$ themselves generate a self-contained Clifford algebra $\mathbf{C}(1,3)$ which is spanned by:
\begin{equation}\label{eq: gen_CLZ}
\begin{split}
g_{0}:& \{I_{2}\otimes I_{2}\otimes I_{2} \} \\
g_{1}:& \{ \sigma^{1}\otimes I_{2}\otimes I_{2}, i\sigma^{2}\otimes I_{2}\otimes \sigma^{k} \} := \{\Gamma^{0}, \Gamma^{k}\}\\
g_{2}:& \{I_{2}\otimes I_{2}\otimes \sigma^{k}, i\sigma^{3}\otimes I_{2}\otimes\sigma^{l}\}:= \{J^{k}, K^{l} \}\\
g_{3}:& \{\sigma^{2}\otimes I_{2}\otimes I_{2}, i\sigma^{1}\otimes I_{2} \otimes \sigma^{k} \} :=\{ \Gamma^{*0}, \Gamma^{*k}\}\\
g_{4}:& \{\sigma^{3}\otimes I_{2}\otimes I_{2}\} : =\{\Gamma^{5}\}
\end{split}
\end{equation}
where all the basis vectors are listed according to their grades $g_{d}.$ This Clifford algebra in literature is referred to as the spacetime algebra. The spacetime algebra itself forms a Lie algebra, with the Lie brackets naturally defined by $[a,b]=a\cdot b-b\cdot a$. Thus from now on $\mathbf{C}(1,3)$ be referred to both as  the Clifford algebra and the Lie algebra. The Lie algebra $\mathbf{C}(1,3)$ can also be obtained as the smallest Lie algebra enveloping the Lorentz algebra and $u(2),$ the Lie algebra of the $U(2)$ group acting on a pair of left and right handed Weyl spinors (i.e. one Dirac fermion). That is to say, when there is only one Dirac fermion, or two Weyl fermions, in each generation, the full symmetry group of the system is generated at least by $\mathbf{C}(1,3)$. This case will be discussed later as the abelian Higgs model derived purely from a gauge theory with the gauge group generated by $\mathbf{C}(1,3)$.  

When there are two Dirac fermions in each generation, it is a bit more complicated to obtain the underlying algebra. Let us first view it as a Lie algebra. The Lie-algebra $u(4)$ defined in Eq.{\ref{eq: gen_U(4)}} and the Lorentz algebra as the sub-algebras together generate a smallest closed algebra with totally 127 basis vectors, two copies for each $\sigma^{\mu}\otimes \sigma^{\nu}\otimes\sigma^{\rho},$ one hermitian and one anti-hermitian, except for $i I_{2}\otimes I_{2}\otimes I_{2}$.  Then, if the derived Lie algebra is considered as a subset of of a geometric algebra, $i I_{2}\otimes I_{2}\otimes I_{2}$ must be added to it. This result is very surprising and intriguing on two aspects. First, as a Lie algebra, both the $u(4)$ and the Lorentz (the spin representation of $so(1,3)$) sub-algebras get complexified, hence both subgroups are now non-compact. This is reminiscent of Weyl's original proposal for unifying electro-magnetism and gravity with "the length-scaling transformation". Algebraically and geometrically there is nothing wrong with the non-compact transformations, such as the Lorentz boosts. But are they all physical? We are unclear yet. Second, the 128 basis vectors actually span a  complexified Clifford algebra of 6 dimensional space, let us denote it as $\mathbf{C}(6)^{C}$. Comparing to the case of only one Dirac fermions in each generation where the geometric algebra is $\mathbf{C}(1,3)$ there seems to be a disparity. A more natural result in the one-Dirac-fermion-per-generation case would be the complexified $\mathbf{C}(1,3)$ so that the Lorentz group would also be complexified. This suggests that the structure of the matter fields is interwoven with the spacetime in a nontrivial way. Or we may speculate that $\mathbf{C}(1,3)$ simply needs to be complexified by hand as that is the truth of nature.

However, let us press on and ignore the physical implications of the full transformation group generated by $\mathbf{C}(6)^{C}$. Let us list these  generators out: 
\begin{equation} \label{eq: gen_nonabelian}
\begin{split}
g_{R}&:=\{ \frac{1}{2}(I_{2}+\sigma^{3})\otimes \sigma^{\mu}\otimes I_{2} \} :=\{ \sigma^{R}\otimes \sigma^{\mu}\otimes I_{2} \}\\
g_{L}&:=\{\frac{1}{2}(I_{2}-\sigma^{3})\otimes \sigma^{\mu}\otimes I_{2} \} :=\{ \sigma^{L}\otimes \sigma^{\mu}\otimes I_{2} \}\\
g_{M}&:=\{ \sigma^{1,2}\otimes \sigma^{\alpha}\otimes I_{2} \} \\
g_{LZ}&:=\{I_{2}\otimes I_{2}\otimes \sigma^{k}, i\sigma^{3}\otimes I_{2}\otimes\sigma^{l}\}:= \{J^{k}, K^{l} \}\\
g_{LM}&:=\{i\sigma^{1,2}\otimes \sigma^{\alpha}\otimes \sigma^{k} \}\\
g_{JK}&:=\{i \sigma^{3}\otimes \sigma^{k}\otimes \sigma^{l}, I_{2}\otimes \sigma^{k}\otimes \sigma^{l}\}\\
g_{C}&:=\{ i\tau \textrm{ for } \tau \in g_{R}, g_{L}, g_{M}, g_{LZ}, g_{LM} \textrm{ and } g_{JK} \}
\end{split}
\end{equation}
When $\mathbf{C}(6)^{C}$ is viewed as the  geometric algebra, the generating basis of the real algebra must contain the four Dirac matrices. The remaining two could be chosen as $\sigma^{3}\otimes \sigma^{1}\otimes I_{2}$ and $\sigma^{3}\otimes \sigma^{2}\otimes I_{2},$ corresponding to a geometric space with the metric:
\begin{equation}
g_{AB}= diag(1,-1,-1,-1,1,1)
\end{equation}
It is noticed that, had the last two signature in the metric be changed to alternative combinations, such as $(-1, -1)$, that would eventually lead to the same geometric algebra $\mathbf{C}(6)^{C}.$ Finding the correct combination of the metric signature in the extra-2 dimensions is a problem. 

Now let us introduce a new symbol $h=\Gamma^{0}$ into the Lagrangian:
\begin{equation} \label {eq: Kinetic}
L=\Psi^{\dagger}h  i \slashed \partial \Psi.
\end{equation}
This is not merely a change of notation, but represents a new concept. The $h$ is commonly referred to as spinor metric. A spinor metric serves the purposes such as defining a norm in the spinor space and making the fermion bilinears real. These constraints lead to a natural choice $h=\Gamma^{0}$ \cite{Crawford}. When these constraints are relaxed the spinor metric does not have to be equal to $\Gamma^{0}$. The spinor metric also transforms differently from the $\Gamma^{\mu} s:$ the Gamma's transform as the generators of the Clifford algebra through similarity transformations, while $h$ transforms as metric through hermitian conjugations, in order to stay hermitian. This difference is not noticed when the transformation is unitary while becoming evident under a non-unitary transformation, e.g. a Lorentz boost. It is appealing to ask about the geometric significance of the norm defined by $h$, but this is not going to be pursued further in this work. Instead, merely the transformation rule of $h$ is exploited. In the following text $h=\Gamma^{0}$ or $h=\gamma^{0}$ is assumed. 

Thus, each of the previously defined 128 generators in $\mathbf{C}(6)^{C}$ generates a transformation $U$ that becomes a symmetry of the Lagrangian with the transformation rule for the spinor metric $h$. So a global transformation that leaves the Lagrangian in Eq.(\ref{eq: Kinetic}) invariant is now expressed as:
\begin{equation}
\Psi \rightarrow U\Psi, \Gamma^{\mu} \rightarrow U\Gamma^{\mu} U^{-1}, h \rightarrow (U^{-1})^{\dagger} h U^{-1}
\end{equation}
with $U$ generated by generators in Eq.(\ref{eq: gen_nonabelian}). This transformation does not spoil the defining property of the Gamma matrices: their anti-commutators are the spacetime metric multiplied by the identity matrix. 

\subsection{Visible and Invisible gauge field components}
Upon being gauged, each generator in $\mathbf{C}(6)^{C}$ is associated with a gauge field component. The gauge field couples to the fermions, and is multiplied to the spinor metric and the Gamma matrices $h\Gamma^{\mu}$. This multiplication does not always produce hermitian product. Thus the average of the Lagrangian and its hermitian conjugate must be taken. As a result, some gauge field components decouple from the fermions in certain spacetime directions. Let us call the decoupled gauge field components (depending on both the index of the group generator and the index of spacetime direction) invisible and remaining coupled visible. Lets us explain this in details. 

When the gauge fields are introduced, as a typical Yang-Mills theory the Lagrangian reads:
 \begin{equation} 
L=\Psi^{\dagger}h  i \slashed \partial \Psi +g\Psi^{\dagger}h  \Gamma^{\mu} A_{\mu} \Psi - \frac{1}{4} \mathbf{tr} F_{\mu \nu} F^{\mu \nu},
\end{equation}
where as usual for each group generator a gauge component is introdueced: $A_{\mu}=A_{h, \mu}^{\alpha,\beta,\gamma}(x)\sigma^{\alpha}\otimes \sigma^{\beta}\otimes \sigma^{\gamma} + i A_{a, \mu}^{\alpha,\beta,\gamma}(x)\sigma^{\alpha}\otimes \sigma^{\beta}\otimes \sigma^{\gamma},$ where the gauge field components $A_{h, \mu}^{\alpha,\beta,\gamma}(x)$ and $A_{a, \mu}^{\alpha,\beta,\gamma}(x)$ are all real, $h$ and $a$ in the subscripts stand for "hermitian" and "anti-hermitian".  In the following text for convenience let us assume that not all the gauge components in the gauge field are excited (up to certain gauge transformations). Instead, only those most relevant for the future calculations, namely those associated to the generators in $g_{R}, g_{L}, g_{M}$ and $g_{LM}$ will be present in the gauge field:
 \begin{equation}
 \begin{split}
A_{\mu}=&A_{R \mu}^{\alpha}(x) \sigma^{R} \otimes \sigma^{\alpha}\otimes I_{2} +A_{L \mu}^{\alpha}(x) \sigma^{L}\otimes \sigma^{\alpha} \otimes I_{2} \\ &+A_{M\mu}^{1, \alpha}(x)\sigma^{1}\otimes \sigma^{\alpha}\otimes I_{2} +A_{M\mu}^{2, \alpha}(x)\sigma^{2}\otimes \sigma^{\alpha}\otimes I_{2} \\  &+iA_{LM\mu}^{1,\alpha, k}(x)\sigma^{1}\otimes\sigma^{\alpha}\otimes \sigma^{k} \\&+iA_{LM\mu}^{2, \alpha, k}(x)\sigma^{2}\otimes\sigma^{\alpha}\otimes \sigma^{k} \end{split}
 \end{equation}
The field strength is defined as usual $F_{\mu \nu}= \partial_{\mu}A_{\nu} -\partial_{\nu}A_{\mu} - ig [ A_{\mu}, A_{\nu} ]$ with the coupling constant $g$ being real. 

Because of the non-hermicity of the product formed by $h\Gamma^{\mu}$ and the group generators, the Lagrangian is modified into:
\begin{equation} \label{eq: Lagrangian}
\begin{split}
L=&\frac{1}{2}\Psi^{\dagger}(h\Gamma^{\mu} i \partial_{\mu} + i \partial_{\mu}h\Gamma^{\mu}) \Psi \\
    &+\frac{1}{2}g\Psi^{\dagger}(h  \Gamma^{\mu} A_{\mu} \Psi + A^{\dagger}_{\mu} h\Gamma^{\mu})  \Psi   \\
    &- \frac{1}{8} \mathbf{tr} (F_{\mu \nu} F^{\mu \nu} +F^{\dagger}_{\mu \nu} F^{\dagger \mu \nu}),
\end{split}
\end{equation} 
where the identity $(h\Gamma^{\mu})^{\dagger}= h\Gamma^{\mu}$ has been used (in fact other gauge invariant choices for making the Yang-Mills curvature term real are also possible. For instance, $\frac{1}{4}\mathbf{tr}hF_{\mu\nu}h^{-1}F^{\dagger \mu\nu}$. The presence of the spinor metric, or any other "metric", is necessary because its transformation rule guarantees the invariance of this term also under non-compact symmetry transformations. If $h$ is chosen to be equal to $I_{8},$ a positive definite metric on the internal space is chosen.) Some curious effect occurs to the gauge field components associated to the generators in $g_{M}$ and $g_{LM}$. Because of non-commutativity between these chirality mixing generators and the product $h\Gamma^{k}$, and the non-compactness of the generators in $g_{LM}$, some gauge field components decouple from the fermions: 
\begin{equation} \label{eq: GaugeCouple}
\begin{split}
\frac{g}{2} \Psi^{\dagger} h\Gamma^{\mu}A_{\mu} \Psi &+h.c. = \frac{g}{2} \Psi^{\dagger} (h\Gamma^{\mu}A_{\mu} + A^{\dagger}_{\mu}h\Gamma^{\mu} )\Psi \\
 =& g\Psi^{\dagger} h\Gamma^{\mu}A_{R\mu}\Psi  +g\Psi^{\dagger} h\Gamma^{\mu}A_{L\mu}\Psi\\
  & + \begin{cases}
 \begin{split}
&\mu= 0 \ \   g \Psi^{\dagger} h\Gamma^{0}A_{M0}^{a,\alpha} \sigma^{a}\otimes\sigma^{\alpha}\otimes I_{2}\Psi   \\
& \mu= k\ \  g \Psi^{\dagger}h\Gamma^{\bar{k}} iA_{LM\bar{k}}^{a,\alpha, \bar{k}}\sigma^{a}\otimes \sigma^{\alpha}\otimes \sigma^{\bar{k}}\Psi
 \end{split}
 \end{cases}\\ 
 := & g \Psi^{\dagger} h\Gamma^{\mu}(A_{keep, \mu} + A_{mix, \mu})\Psi
  \end{split}
 \end{equation}
where the summation over the latin index $a$ runs over $1,2,$  the bar over the index $k$ indicates no summation performed, and $A_{\mu} = A_{keep, \mu} + A_{mix, \mu}$ consists of two distinct sets of gauge components, namely chirality keeping and mixing. A closer look gives that the spatial chirality mixing imaginary gauge components are equivalent to temporal chirality mixing real gauge components as far as only the fermion interaction is concerned:  
\begin{equation}
\begin{split}
g \Psi^{\dagger}h\Gamma^{\bar{k}}( iA_{LM\bar{k}}^{1,\alpha, \bar{k}}\sigma^{1}\otimes \sigma^{\alpha}\otimes \sigma^{\bar{k}}+iA_{LM\bar{k}}^{2,\alpha,\bar{k}}\sigma^{2}\otimes \sigma^{\alpha}\otimes \sigma^{\bar{k}} )\Psi &\\
=g \Psi^{\dagger} h\Gamma^{0} (A_{LM \bar{k}}^{1,\alpha,\bar{k}} \sigma^{2}\otimes \sigma^{\alpha}\otimes I_{2} -A_{LM\bar{k}}^{2,\alpha,\bar{k}} \sigma^{1}\otimes \sigma^{\alpha}\otimes I_{2})\Psi &
\end{split}
\end{equation}
The calculation is performed with the choice that $h\Gamma^{k}= -\sigma^{3}\otimes I_{2}\otimes \sigma^{k}$. But clearly the equality holds true against any transformation on the Clifford algebra. In the standard model, the Higgs field is a $SU(2)_{L}$ doublet consisting of 4 real components. The Yukawa terms involving only the fermions and the Higgs field can be rewritten precisely as gauge field in the temporal direction with the chirality mixing generators in $g_{M}$ subjected to some constraint. This suggests the identification of chirality mixing gauge components collectively as the Higgs field. 
\section{Problems with the Yukawa terms and Resolutions}
The Yukawa terms involving one generation of fermions in the standard model are:
\begin{equation} \label{eq: YSM_1}
L_{Y}=m_{u}\epsilon^{ab}\Psi^{\dagger}_{L a} \Phi_{b}^{\dagger}\Psi_{uR} +m_{d}\Psi^{\dagger}_{L}\cdot \Phi \Psi_{dR} + h.c.
\end{equation}
where $\epsilon^{ab}$ is anti-symmetric in $a,b \in \{0, 1 \}$, while $m_{u,d}$ are proportional to the fermions' masses, for instance, the masses of up and down quarks, or neutrinos and the charged leptons respectively. Let's denote the Higgs doublet as:
\begin{equation} \label{eq: YSM_2}
\Phi(x)=\begin{pmatrix}
\phi_{1}+i\phi_{2}\\
\phi_{3}-i\phi_{4}
\end{pmatrix}
\end{equation}
Then the Yukawa terms can be written as:
\begin{equation} \label{eq: Yukawa}
\begin{split}
L_{Y}&= \Psi^{\dagger}h\Gamma^{0} [m_{a}( \phi_{1} \sigma^{2}\otimes \sigma^{2}\otimes I_{2} +\phi_{2} \sigma^{2}\otimes \sigma^{1}\otimes I_{2}\\
& +\phi_{3} \sigma^{1}\otimes I_{2}\otimes I_{2} +\phi_{4} \sigma^{2}\otimes \sigma^{3}\otimes I_{2})  \\
         &+m_{b}(-\phi_{1} \sigma^{1}\otimes \sigma^{1}\otimes I_{2} +\phi_{2} \sigma^{1}\otimes \sigma^{2}\otimes I_{2}\\
        &  +\phi_{3} \sigma^{1}\otimes \sigma^{3}\otimes I_{2} +\phi_{4} \sigma^{2}\otimes I_{2}\otimes I_{2}) ]\Psi ,
         \end{split}
         \end{equation}
where $m_{a}=(m_{1}+m_{2})/2$ and $m_{b}=(m_{1}-m_{2})/2.$\\

When Eq.(\ref{eq: GaugeCouple}) is compared to Eq.(\ref{eq: Yukawa}), two crucial differences are noticed: 1) in Eq.(\ref{eq: GaugeCouple}) every fermion doublet (Dirac fermions) must couple to the to-be-identified Higgs field (the chirality mixing gauge components $h\Gamma^{\mu}A_{mix,\mu}$) with the same strength, i.e. the gauge coupling constant $g$. But in Eq.(\ref{eq: Yukawa}) in the real standard model, fermions from different generations couple to the Higgs field with different strength (mass ratios). 2) in Eq.(\ref{eq: GaugeCouple}) the to-be-identified Higgs field as gauge field components transform as a vector. However, in Eq.(\ref{eq: Yukawa}) in the real standard model it transforms as a scalar.  These two conflicts need to be resolved. The resolutions are found to rely on the Lorentz transformations.
\subsection{Two types of Lorentz transformations}
Let's first address the problem of the varying coupling strength for different generations of fermions, i.e. the mass ratio problem. That is to say, the chirality mixing gauge components must couple to different fermions with different mass parameters. The resolution turns out to rely on a special arrangement with the Lorentz transformations and assignment of  Gamma matrices for each generation of fermions. The chirality mixing generators in $g_{M}$ and $g_{LM}$ do not commute with the Lorentz transformations. The associated gauge field components are mixed under a rotation and get rescaled under a boost. But for different generations of fermions these field components need not to be rescaled and mixed in the same way, despite a "common" Lorentz transformation is performed. This can be understood as, the actual transformation exerted on each generation of fermions is a representation of an abstract Lorentz transformation, with unique parameters characterising the fermions. Put in a different way, this is like a non-abelian counterpart of the electro-magnetic transformation--- fermions with different charges are rotated differently under the same abstract $U(1)$ transformation. To incorporate with this special arrangement, different copies of Gamma matrices are assigned to couple to different generations of fermions at first. Then they are brought to the same copy by an abstract Lorentz transformation. Let us get this idea clarified. 

The Lagrangian defined in Eq.(\ref{eq: Lagrangian}) admits two types of global Lorentz transformations: one ordinary transformation in which the spinors, the spacetime points and the tangent vectors are all transformed simultaneously, but in which the Gamma matrices are left invariant. This type will be referred to as external Lorentz transformations. Following Peskin and Schroeder let us denote the spacetime representation of the proper Lorentz group by $\Lambda \in SO(1,3)$ and its spin representation by $\Lambda_{\frac{1}{2}}$. An external Lorentz transformation acts on the relevant component of the Lagrangian according to:
\begin{equation} \label{eq: ex-transformation}
\begin{split}
&x \rightarrow x'=\Lambda x,  \partial_{x} \rightarrow \partial_{x}= \Lambda^{-1} \partial_x', \Gamma^{\mu}\rightarrow \Gamma^{\mu}\\
&\Psi(x) \rightarrow \Lambda_{\frac{1}{2}} \otimes \Lambda \cdot \Psi(x)=\Lambda_{\frac{1}{2}} \Psi(x')\\
&A_{\mu}(x) \rightarrow (\Lambda^{-1})^{\nu} _{\mu}\cdot \Lambda_{\frac{1}{2}}\otimes \Lambda \cdot A_{\nu}(x)\cdot\Lambda^{-1}_{\frac{1}{2}}\otimes \Lambda^{-1}\\ &=(\Lambda^{-1})^{\nu}_{\mu}\Lambda_{\frac{1}{2}}A_{\nu}(x')\Lambda^{-1}_{\frac{1}{2}}.
\end{split}
\end{equation}
This is the usual way of defining the Lorentz transformations of the Lagrangian, which simultaneously act on both the internal gauge space and the external Minkowskian space. 

The second type of Lorentz transformations act only on the internal gauge space, as well as on the Gamma matrices. They will referred to as internal Lorentz transformations. In other words, the internal Lorentz transformations can be seen as merely producing a new matrix representation of the Clifford algebra. These transformations act according to the following rule:
\begin{equation} \label{eq: in-transformation}
\begin{split}
&x \rightarrow  x,  \partial_{x} \rightarrow   \partial_{x} \\
&\Psi(x) \rightarrow \Lambda_{\frac{1}{2}}\Psi(x) \\
&h\rightarrow (\Lambda^{-1}_{\frac{1}{2}} )^{\dagger} h \Lambda^{-1}_{\frac{1}{2}}=h,   \Gamma^{\mu}\rightarrow \Lambda_{\frac{1}{2}} \Gamma^{\mu} \Lambda^{-1}_{\frac{1}{2}}= (\Lambda^{-1})^{\mu}_{\nu}\Gamma^{\nu} \\
&A_{\mu}(x) \rightarrow \Lambda_{\frac{1}{2}} A_{\mu}(x) \Lambda^{-1}_{\frac{1}{2}}.
\end{split}
\end{equation}
It is noticed that when $h=\Gamma^{0}$, under a Lorentz transformation $h$ remains the same, while the $\Gamma$ matrices are transformed like a vector.\\

A closer look at the internal Lorentz transformations inspires us to look at the external Lorentz transformations in an alternative way: the external Lorentz transformations can be seen as internal ones combined with a spacetime coordinate-transformation which coincides with the spacetime representation of the Lorentz transformation. In principle, the internal transformation and the spacetime coordiante-transformation can be independent from each other and thus performed separately. If the discussion is constrained to global Lorentz transformations, it is seen that the spacetime transformation causes the transformation on $x_{\mu}$ and the measure of the integration, which only reparameterises the spacetime dependence of all the fields (as well as the curvature terms). The internal Lorentz transformations, on the other hand, act on the spinors, the spinor metric, the Gamma matrices and the gauge field. The spinor metric $h=\Gamma^{0}$ is invariant under the Lorentz transformation. The transformed Gamma matrices $\Gamma^{\mu}\Rightarrow \tilde{\Gamma}^{\mu}= \Lambda_{\frac{1}{2}}\Gamma^{\mu}\Lambda^{-1}_{\frac{1}{2}} = \Lambda_{\nu}^{-1\mu}\Gamma^{\nu}$ and the the gauge field $A_{\mu}\Rightarrow \tilde{A}_{\mu} =  \Lambda_{\frac{1}{2}}A_{\mu}\Lambda^{-1}_{\frac{1}{2}}.$ This can be equivalently seen as
\begin{equation}
h\Rightarrow h, \Gamma^{\mu}\Rightarrow \Gamma^{\mu}, \partial_{\mu} \Rightarrow  \Lambda_{\mu}^{-1\nu} \partial_{\nu}, A_{\mu}\Rightarrow  \Lambda_{\mu}^{-1\nu} \Lambda_{\frac{1}{2}}A_{\nu}\Lambda^{-1}_{\frac{1}{2}}
\end{equation}
At this stage one may ask, are transformations other than the Lorentz transformations also accompanied by a transformation on the spacetime? Since the internal symmetry transformation and the spacetime transformation can be performed separately, it seems that this question becomes irrelevant. 

When the transformations are made local, the above decomposition of an external Lorentz transformation into an internal one followed by a spacetime transformation can be done in a similar way, except that extra attention must be paid to the order between the partial derivative and the the Gamma matrices in the kinetic part of the Lagrangian. A natural modification to resolve this issue is to define the kinetic term to be:
\begin{equation}
\begin{split}
\Psi^{\dagger}h\Gamma^{\mu} i \partial_{\mu}\Psi \rightarrow& \frac{1}{2}[\Psi^{\dagger}h\Gamma^{\mu} i \partial_{\mu}\Psi + (\Psi^{\dagger}h\Gamma^{\mu} i \partial_{\mu}\Psi)^{\dagger} ]\\
=&\frac{1}{2}\Psi^{\dagger}[h\Gamma^{\mu} i \partial_{\mu}+ i\partial_{\mu}h\Gamma^{\mu} ]\Psi 
\end{split}
\end{equation}
as has been guessed earlier. 

\subsection{Resolution to the mass-ratio problem}
Let's now conceive a new theory where the Lagrangian defined in Eq.(\ref{eq: Lagrangian}) is modified into:
\begin{equation} \label{eq: Lo}
\begin{split}
L_{o}=& \sum_{m} \frac{1}{2}[\Psi_{m}^{\dagger}h\Gamma_{m}^{\mu}  (i \partial_{\mu} + g  A_{\mu})\Psi_{m} +h.c.]  \\
           &- \frac{1}{8} \mathbf{tr} [F_{\mu \nu} F^{\mu \nu}+F^{\dagger}_{\mu \nu} F^{\dagger \mu \nu}]
           \end{split}
 \end{equation}
 where $\Gamma^{\mu}_{m}=\Lambda^{-1}_{\frac{1}{2}}(m) \Gamma^{\mu} \Lambda_{\frac{1}{2}}(m) =\Lambda^{\mu}_{\nu}(m) \Gamma^{\nu}$, i.e. $\Gamma^{\mu}_{m}s$ are obtained from $\Gamma^{\mu}s$ by a Lorentz transformation with parameters collectively denoted by $m$ which contains information about the masses of the fermion doublets $\Psi_{m}.$ Let's abstractly denote the set of these Lorentz transformations $\Lambda_{\frac{1}{2}}(m)$ with varying $m$ by $\Lambda_{\frac{1}{2}M}$. Each  $\Lambda_{\frac{1}{2}}(m)$ can be seen as a representation of $\Lambda_{\frac{1}{2}M}$ with parameters $m$, in the same sense as the $U(1)_{EM}=\exp(i q\phi)$ phase of a charge-$q$ particle living in a particular representation of the same $U(1)$. This modification does not affect the general transformations $U$ discussed previously, be it global or local. Let's specify the symmetry transformations of the Lagrangian $L_{o}$ to show this point explicitly.\\

1) Global transformations of $L_{o}$ including the external Lorentz transformations.
 
It is clear that under the following transformation the Lagrangian $L_{o}$ stays invariant. 
\begin{equation}
\begin{split}
&\Psi_{m} \rightarrow U\Psi_{m}, A_{\mu}\rightarrow UA_{\mu}U^{-1},\\
&\Gamma_{m}^{\mu} \rightarrow U\Gamma_{m}^{\mu} U^{-1}, h \rightarrow (U^{-1})^{\dagger} h U^{-1}.
\end{split}
\end{equation}
Let's have a closer look at the transformed Gamma matrices:
\begin{equation} \label{eq: GlobalU_Lo}
\begin{split}
\tilde{\Gamma}_{m}^{\mu} & = U\Gamma_{m}^{\mu} U^{-1}\\
                                  & =U \Lambda^{-1}_{\frac{1}{2}}(m) \Gamma^{\mu} \Lambda_{\frac{1}{2}}(m)U^{-1} \\
                                  & = U \Lambda^{-1}_{\frac{1}{2}}(m) U^{-1} U \Gamma^{\mu}U^{-1} U  \Lambda_{\frac{1}{2}}(m) U^{-1} \\                               
                                  & := \tilde{\Lambda}^{-1}_{\frac{1}{2}}(m) \tilde{\Gamma}^{\mu} \tilde{\Lambda}_{\frac{1}{2}}(m)\\
                                  &= \Lambda_{\nu}^{\mu}(m) \tilde{\Gamma}^{\mu}
                               \end{split}
\end{equation}
The result can also be seen more directly: $\tilde{\Gamma}_{m}^{\mu}  = U\Gamma_{m}^{\mu} U^{-1}=\Lambda_{\nu}^{\mu}(m) U\Gamma^{\mu}U^{-1}=\Lambda_{\nu}^{\mu}(m) \tilde{\Gamma}^{\mu}.$ This simply means that the similarity transformation on the whole Clifford algebra does not alter the mass parameters contained in $\Gamma^{m}$, a necessary condition that follows from that fermions must have well-defined masses. Transformation rule of the gauge field can be obtained by following the transformation of any one generation of the fermions, ending up with the same result. Finally, if the transformation is an external Lorentz transformation, it can be equivalently decomposed into an internal one and a corresponding coordinate transformation, both are symmetries of $L_{o}$ and can be performed consequtively. Therefore, the global symmetry transformations of $L_{o}$ are consistently defined.\\

2) The gauge transformation $U(x)$ of $L_{o}$ including the external Lorentz transformations.\\
The above global symmetry of Lagrangian $L_{o}$ can be made local according to: 
\begin{equation}
\begin{split}
&\Psi(x) \rightarrow U(x)\Psi(x), \\
&\Gamma_{m}^{\mu} \rightarrow U(x)\Gamma_{m}^{\mu} U^{\dagger}(x), h \rightarrow (U^{-1})^{\dagger}(x) h U^{-1}(x),\\
&A_{\mu}\rightarrow U(x)A_{\mu}U^{\dagger}(x)-i/g \partial_{\mu} U(x) U^{\dagger}(x).
\end{split}
\end{equation}
Consistency can be shown in the same way as in the global case, except for allowing $\tilde{\Gamma}_{m}^{\mu}$, $h$ and $U(x)$ now space-time dependent, i.e. a local similarity transformation on the Clifford algebra is dealt with. In case of local external Lorentz transformation, the part involving the spacetime coordinate transformation needs to be interpreted in terms of geometric algebra. Details can be found in \cite{DL}. Gauge transformations of $L_{o}$ are thus consistently defined. \\

So the Lagrangian $L_{o}$ defines a consistent interacting theory. How do we understand the extra-ordinary feature, that different $\Gamma_{m}^{\mu}$ are assigned to couple to different fermions? The plain fact that these various copies of $\Gamma_{m}^{\mu}$ despite being different are all valid generating basis of the space-time algebra ------ the geometric algebra of our spacetime. None should be discriminated from the others. Thus each copy provides a valid reference to the spacetime for the matter fields. They are different meaning that different matter fields reference to the spacetime in different manners. This should be and can be allowed by a valid theory. In the standard model, these various copies of Gamma matrices are brought (transformed) to the same copy, just to serve us the convenience by providing a simpler perspective to look at the interaction between matter and the spacetime, which filters the true complexity away. Perhaps this is also the right opportunity to discuss the possible connection between the Higgs field and the gravity. The chirality mixing transformations do not commute with the Lorentz boost transformations. Thus they need to be considered together with the Lorentz transformations in a larger group, to form a consistent theory. The gauged Lorentz transformations describe partially the gravity, and the chirality mixing gauge field components are collectively seen as the Higgs field. This suggests that the Higgs field and the gravitational field are interwoven together. But in the standard model gravity is not treated while the Higgs field must be present. This seems to suggest that we should look at the Higgs field as some constrained excitation and no local chirality mixing transformations should be permitted (otherwise, they can generate equivalent local Lorentz transformations). \\

Starting with the Lagrangian $L_{o}$ let us perform the internal Lorentz transformation $\Lambda_{\frac{1}{2}M}$ (to each generation the corresponding representation $\Lambda_{\frac{1}{2}}(m)$ is applied.) Two aspects of the effect are immediately observed: 1, to each generation of fermion the associated $\Gamma_{m}^{\mu}$ are transformed back to the common copy $\Gamma^{\mu}$; 2, the gauge field $A_{\mu}$ coupled to this generation is transformed into : $A_{\mu} \rightarrow \tilde{A}_{\mu}=\Lambda_{\frac{1}{2}}(m) A_{\mu}(x) \Lambda^{-1}_{\frac{1}{2}}(m)$.  The transformed gauge field shows desired features: the chirality mixing gauge components (to be identified as the Higgs field) get scaled, while the chirality keeping components stay intact (because the chirality keeping generators commute with the Lorentz transformations). The scaling of the chirality mixing gauge field components can be equally viewed as the scaling of the coupling strength between the gauge field components and the fermions. In this way, the various coupling constants $m$ between the Higgs and each generation of fermions are understood. 

Let's demonstrate this point by considering only the chirality mixing generators in $g_{4M}$ and a boost transformation $\exp(\hat{\theta}_{k}\sigma^{3}\otimes I_{2}\otimes \sigma^{k})$ followed by a rotation (which commutes with the generators in $g_{4M}$ and only helps with transforming $\Gamma_{m}^{\mu}$). The relevant term is transformed like this:
\begin{equation}
\begin{split}
&\Psi^{\dagger} h\Gamma_{m}^{\mu}A_{\mu}\Psi \Rightarrow \tilde{\Psi}^{\dagger} h\Gamma^{\mu} \tilde{A}_{\mu}\tilde{\Psi}\\
&=\tilde{\Psi}^{\dagger}h\Gamma^{\mu} A_{keep, \mu}\tilde{\Psi} + \tilde{\Psi}^{\dagger}h \Gamma^{\mu} \cosh \theta A_{mix,\mu}\tilde{\Psi} +\\
&\    \ +\begin{cases}
\begin{split}
& \mu=0 \    \   \tilde{\Psi}^{\dagger} h\Gamma^{l} \hat{\theta}_{l}\sinh \theta A_{mix, 0} ^{b} \tau^{b} \tilde{\Psi} \\
& \mu=l \    \   \tilde{\Psi}^{\dagger} h\Gamma^{0} \hat{\theta}_{l}\sinh \theta A_{mix, l}^{b}\tau^{b} \tilde{\Psi} \\
& \mu=j \neq l \    \   \tilde{\Psi}^{\dagger} h\Gamma^{k} \epsilon^{j l k}\hat{\theta}_{l}\sinh \theta A_{mix, j}^{b} i Z\cdot \tau^{b} \tilde{\Psi}, 
\end{split}
\end{cases}\end{split}
\end{equation}
where the operator $Z=\sigma^{3}\otimes I_{2}\otimes I_{2}.$ Eventually after the internal Lorentz boost $\Lambda^{-1}_{M}$ the visible chirality mixing gauge components becomes 
\begin{equation} \label{eq: in-boost}
\tilde {A}_{v, mix}=\cosh \theta A_{mix,0} + \sinh \theta \hat{\theta}_{k} A_{mix, k}
\end{equation} 
$\tilde{A}_{v, mix}$ is defined in the sense that $\tilde{\Psi}^{\dagger} h\Gamma_{m}^{\mu}A_{mix, \mu}\tilde{\Psi} =\tilde{\Psi}^{\dagger}A_{v,mix}\tilde{\Psi}.$ It is scaled in a non-trivial way compared to that before the Lorentz transformation. By adjusting the parameters $(\theta, \theta^{1}, \theta^{2}, \theta^{3})$ (later it will be seen that parameters associated to Lorentz rotations are also necessary ) various coupling constants for the fermions and the Higgs field can be realised. Thus the mass information was originally hidden in the Gamma matrices $\Gamma_{m}$ as present in the Lagrangian $L_{o}$. Only after the internal boost $\Lambda^{-1}_{M}$ the mass ratio parameters are revealed, as explicitly seen in the transformed Lagrangian:
\begin{equation} \label{eq: Lagrangian_t}
\begin{split}
L_{t}=& \sum_{m} \frac{1}{2}[\Psi_{m}^{\dagger}h\Gamma^{\mu}( i\partial_{\mu} + g A_{keep,\mu})\Psi_{m} +h.c]\\
&+ \sum_{m} \Psi_{m}^{\dagger} m\cdot A_{v,mix}\Psi_{m}\\
&-\frac{1}{8} \mathbf{tr} (F_{\mu \nu} F^{ \mu \nu} +F^{\dagger}_{\mu \nu} F^{\dagger \mu \nu} )
\end{split}
 \end{equation}
Note that the last term, the contribution from the field strength, is invariant under the specified internal boost transformation, no matter which representation of the transformation was applied to the gauge field. This also reminds us that, for calculating $F_{\mu \nu}$ knowing only the visible part of the chirality mixing gauge components $A_{v,mix}$ is not enough. $L_{t}$ must be always supplemented with a book-keeping of the full chirality mixing gauge components $A_{mix, \mu},$ or equivalently $L_{o}$ must be known. In other words, $L_{o}$ should be considered as the fundamental theory while $L_{t}$ is adopted to suit our cognitive habit shaped by the concept of mass. \\

However, inconsistency seems to arise if a further gauge transformation is naively performed (e.g. a chirality mixing transformation) to $L_{t}$. The coupling constants between the chirality mixing gauge components and the fermions would change, making the "coupling constants" not constants any more. This observation is actually not true. The reason is that to perform a transformation  $T$ (be it global or local) to $L_{t}$, we must first go back to $L_{o}$ by the inverse internal Lorentz boost $\Lambda^{-1}_{\frac{1}{2}M}$. Then the transformation $T$ is performed to $L_{o}$ to obtain $\tilde{L}_{o}.$ After that $\Lambda_{\frac{1}{2}M}$ is performed to the transformed $\tilde{L}_{o}$, to get the final transformed Lagrangian $\tilde{L}_{t}.$ In other words, any transformation $T$ to be applied, can either be directly applied to $L_{o}$, or applied to $L_{t}$ in the conjugated form of $\Lambda_{\frac{1}{2}M,2}\cdot T \cdot \Lambda^{-1}_{\frac{1}{2}M,1}.$ This defines a consistent gauge theory $L_{t}$.  Note that in the sequence of the three transformations, the last Lorentz transformation is actually working in the $T-$ transformed representation of the Clifford algebra, i.e. $\Lambda_{\frac{1}{2}M,2} = T\Lambda_{\frac{1}{2}M,1}T^{-1} $. With this understanding it can be shown that the sequence $\Lambda_{\frac{1}{2}M,2}\cdot T \cdot \Lambda^{-1}_{\frac{1}{2}M,1}$ has the same effect of $T$ on $L_{t}$ for both $\Gamma^{\mu}$ and the gauge field components (including both chirality keeping and mixing ones). However, the transformation on the spinor metric must be taken special care of. It is transformed into $h \rightarrow (\Lambda^{-1}_{\frac{1}{2}M,2})^{\dagger}\cdot (T^{-1})^{\dagger} \cdot \Lambda^{\dagger}_{\frac{1}{2}M,1} h \Lambda_{\frac{1}{2}M,1}\cdot T^{-1} \cdot \Lambda^{-1}_{\frac{1}{2}M,2}=(T^{-1})^{\dagger}hT^{-1}.$ Thus $h$ is not always invariant under a symmetry transformation. In principle it could even be different for different fermions. This does not ruin the consistency of the theory verified earlier. It is due to the observer's preference, the special perspective is chosen so that $h$ is the same for all fermions and some certain symmetry transformations are suppressed.\\

Let us give an example to illustrate the above discussion by performing a global internal Lorentz transformation $L_{\frac{1}{2}}$ to $L_{t}$:
\begin{equation}
\begin{split}
\tilde{h} =& h\\
\tilde{\Gamma}^{\mu}=&\Lambda_{\frac{1}{2}M,2} L_{\frac{1}{2}} \Lambda^{-1}_{\frac{1}{2}M,1} \Gamma^{\mu}\Lambda_{\frac{1}{2}M,1}L_{\frac{1}{2}} \Lambda^{-1}_{\frac{1}{2}M,2}\\
= &\Lambda_{\frac{1}{2}M,2} L_{\frac{1}{2}}\Gamma_{M}^{\mu}L_{\frac{1}{2}}\Lambda^{-1}_{\frac{1}{2}M,2}\\
= &\tilde{\Lambda}_{\frac{1}{2}M} \tilde{\Gamma}_{M}^{\mu} \tilde{\Lambda}^{-1}_{\frac{1}{2}M}=  \tilde{\Gamma}^{\mu} =  L^{-1 \mu}_{\nu} \Gamma^{\nu}\\
\tilde{A}_{\mu}(x) =& \Lambda_{\frac{1}{2}M,2} L_{\frac{1}{2}} \Lambda^{-1}_{\frac{1}{2}M,1} A_{\mu}(x)\Lambda_{\frac{1}{2}M,1}L_{\frac{1}{2}}\Lambda^{-1}_{\frac{1}{2}M,2}\\
= &  L_{\frac{1}{2}}  A_{\mu}  L^{-1}_{\frac{1}{2}}, 
\end{split}
\end{equation}
where $\tilde{\Lambda}_{\frac{1}{2}M}= L_{\frac{1}{2}} \Lambda_{\frac{1}{2}M}L^{-1}_{\frac{1}{2}}$. This then gives a global external Lorentz transformation
\begin{equation}
\begin{split}
\tilde{x}&=L\cdot x:=x', \tilde{\partial}= L^{-1\nu}_{\mu'}\partial_{\nu}:=\partial_{\mu'}, \\
\tilde{h} &= h,  \tilde{\Gamma}^{\mu'}= \Gamma^{\mu},\\
\tilde{A}_{\mu'}(x') &=L^{-1\nu}_{\mu'} \tilde{\Lambda}_{\frac{1}{2}M} L_{\frac{1}{2}} \Lambda^{-1}_{\frac{1}{2}M} A_{\nu}(x')\Lambda_{\frac{1}{2}M}L_{\frac{1}{2}}\tilde{\Lambda}^{-1}_{\frac{1}{2}M},\\
&= L^{-1\nu}_{\mu'} L_{\frac{1}{2}}  A_{\nu} (L\cdot x) L^{-1}_{\frac{1}{2}}.
\end{split}
\end{equation}
When $A_{\mu}$ is restricted to the gauge field components associated to the generators in $g_{L}$ and $g_{R}$ this is exactly the usual Lorentz transformation.
\subsection{The resolution to the scalar-appearance problem of the Higgs field}
Since the chirality keeping generators in $g_{R}$ and $g_{L}$ commute with the Lorentz transformation $L_{\frac{1}{2}}$, it is just as if the Gamma matrices stay the same and the the corresponding gauge field components are transforming like a vector; the chirality mixing generators on the other hand, in general do not commute with the Lorentz transformations and exhibits interesting features. Under a certain condition, they collectively behave as a scalar. Let's explain this point now.  

First, it can be shown that a generic global transformation does not alter the visibility of any gauge components. Under such a transformation $G$, the Yukawa term is transformed as follows:
\begin{equation}
\begin{split}
\Psi^{\dagger} h \Gamma^{\mu} A_{\mu} \Psi& \Rightarrow \tilde{\Psi}^{\dagger} \tilde{h} \tilde{\Gamma}^{\mu} \tilde{A}_{\mu} \tilde{\Psi} \\
 &= \Psi^{\dagger}G^{\dagger}(G^{-1})^{\dagger}h G^{-1}  G\Gamma^{\mu}G^{-1} G A_{\mu}G^{-1}G \Psi\\
& =\tilde{\Psi}^{\dagger}  (G^{-1})^{\dagger}h \Gamma^{\mu} A_{\mu} G^{-1}\tilde{\Psi}
\end{split}
\end{equation}
When the gauge component is visible $(h \Gamma^{\mu} A_{\mu} )^{\dagger} = h \Gamma^{\mu} A_{\mu} $ holds,  thus $[(G^{-1} )^{\dagger}h \Gamma^{\mu} A_{\mu}  G^{-1}]^{\dagger}= (G^{-1} )^{\dagger}h \Gamma^{\mu} A_{\mu}  G^{-1}$, meaning that  $\tilde{A}_{\mu}$ is visible, in the $G-$ transformed basis of the Clifford algebra. When the gauge component is invisible, it can be similarly shown that $\tilde{A}_{\mu}$ is invisible either. A particular case is that $G=\Lambda_{\frac{1}{2}}$ an internal Lorentz transformation, followed by an arbitrary coordinate redefinition. So an external Lorentz transformation particularly does not alter the visibility of the gauge components. 

Second, let us show that in the special case the visible part of the chirality mixing gauge components sum up to 
\begin{equation}\label{eq: Constraint_Higgs}
A_{v,mix}=h\Gamma^{\mu}A_{mix, \mu} = \phi^{1,\alpha} \sigma^{1}\otimes\sigma^{\alpha} \otimes I_{2} + \phi^{2,\alpha} \sigma^{2}\otimes \sigma^{\alpha}\otimes I_{2}
\end{equation}
then these gauge components collectively form an invariant object under the external Lorentz transformations, i.e. behaving as a scalar. 

This can be verified rather straightforwardly by considering the internal Lorentz boosts and rotations separately. The transformed visible chirality gauge components are
 \begin{equation}
 (\Lambda^{-1}_{\frac{1}{2}} )^{\dagger}h \Gamma^{\mu} A_{mix, \mu}  \Lambda^{-1}_{\frac{1}{2}} = h \Gamma^{\mu} A_{mix, \mu} 
 \end{equation}
The equality holds true because, when it is a boost, $(\Lambda^{-1}_{\frac{1}{2}} )^{\dagger}=\Lambda^{-1}_{\frac{1}{2}} $ and since the generator $i\sigma^{3}\otimes I_{2}\otimes \sigma^{k}$ anti-commutes with $\sigma^{1,2}\otimes \sigma^{\alpha}\otimes I_{2}$, $\Lambda^{-1}_{\frac{1}{2}}$ becomes $\Lambda_{\frac{1}{2}}$ when moved from the left side of $h \Gamma^{\mu} A_{mix, \mu} $  to its right and cancels with the factor already being there to the right; when it is a rotation, $(\Lambda^{-1}_{\frac{1}{2}} )^{\dagger}=\Lambda_{\frac{1}{2}} $ and the generator commutes with $\sigma^{1,2}\otimes \sigma^{\alpha}\otimes I_{2}$ then the same cancellation occurs when it is moved from the left side of $h \Gamma^{\mu} A_{mix, \mu} $ to the right. So the term $h\Gamma^{\mu}A_{mix, \mu}$ is invariant under an internal Lorentz transformation.

Based on the previous discussion, when a Lorentz transformation is performed to $L_{t},$  each term associated to a particular generation of fermions in the Lagrangian actually goes through three Lorentz transformations $\Lambda^{-1}_{\frac{1}{2}}(m)$, then $L_{\frac{1}{2}}$, and at last $L_{\frac{1}{2}} \Lambda_{\frac{1}{2}}(m) L_{\frac{1}{2}}^{-1}.$ These  altogether is equivalent to $L_{\frac{1}{2}}$ alone. Let the internal Lorentz transformation $L_{\frac{1}{2}}$ be followed by a corresponding spacetime transformation $L$, which would combine to form an external Lorentz transformation. This then shows that $h\Gamma^{\mu}A_{mix, \mu}$ behaves just like a scalar under the external Lorentz transformation, provided that the constraint in Eq.(\ref{eq: Constraint_Higgs}) is satisfied. 

Thus the mass-ratio problem and the scalar-appearance problem, are both resolved in the case of two Dirac fermions in each generation. It is straightforward to check that the same arguments hold in the case of only one Dirac fermion in each generation. The former leads to the $SU(2)_{L}\times U(1)_{Y}$ non-abelian Higgs model as in the standard model, while the latter leads to the abelian Higgs model. In the following details of these two cases are presented. 

\section{Reconsctruct the abelian and non-abelian Higgs models}

There are two well-known models with the Higgs mechanism. The most successful one is the electro-weak theory with a $SU{2}_{L}$ complex Higgs doublet in the standard model. The other is a $U(1)$ abelian Higgs theory which may be related to the theory of superconductors at low temperature. In this section both models will be derived starting with the corresponding full theory described by $L_{o}$ in Eq. (\ref{eq: Lo}). 

\subsection{The $SU(2)_{L}$ Higgs theory---towards the standard model}

The non-abelian Higgs model is the Weinberg-Salam model with a left-handed $SU(2)_{L}$ Higgs doublet. In the following we are largely following chapter 20.2 of the book by Peskin and Schroeder \cite{PS}. In the standard model the Yukawa terms with the Higgs field coupled to one generation of fermions can be written as:
\begin{equation} \label{eq: Higgs_SM}
\begin{split}
L_{Y}=&m_{d} Q_{L}^{\dagger}\Phi d_{R} +m_{u}Q_{L}^{\dagger} i\sigma^{2}\Phi^{*} u_{R}  + h.c.\\
         =&\Psi^{\dagger} (m_{a}\Phi_{a} + m_{b}\Phi_{b} ) \Psi,
\end{split}
\end{equation}
where $\Psi$ consists of the left and right-handed components of both the up and down quarks (could also be any other generation of quarks or leptons), and $\Phi$ is the complex Higgs doublet:
\begin{equation} \label{eq: Psi}
\Psi=\begin{pmatrix}
u_{R}\\
d_{R}\\
u_{L}\\
d_{L}
\end{pmatrix}
\    \ Q_{L}=\begin{pmatrix}
u_{L}\\
d_{L}
\end{pmatrix}\  \ and \  \  \Phi=\begin{pmatrix}
\phi_{1} + i \phi_{2}\\
\phi_{3} - i \phi_{4}\\
\end{pmatrix}
\end{equation}
and 
\begin{equation} \label{eq: Config_Higgs}
\begin{split}
m_{a}=&\frac{1}{2}(m_{u}+m_{d}), \           \ m_{b}=\frac{1}{2}(m_{u}-m_{d})\\
\Phi_{a}=&\phi_{1} \sigma^{2}\otimes \sigma^{2}\otimes I_{2} +\phi_{2} \sigma^{2}\otimes \sigma^{1}\otimes I_{2}\\
             & +\phi_{3} \sigma^{1}\otimes I_{2}\otimes I_{2} +\phi_{4} \sigma^{2}\otimes \sigma^{3}\otimes I_{2}\\
\Phi_{b}=&-\phi_{1} \sigma^{1}\otimes \sigma^{1}\otimes I_{2} +\phi_{2} \sigma^{1}\otimes \sigma^{2}\otimes I_{2}\\
              &+\phi_{3} \sigma^{1}\otimes \sigma^{3}\otimes I_{2} +\phi_{4} \sigma^{2}\otimes I_{2}\otimes I_{2}
               \end{split}
\end{equation}
with $m_{u}, m_{d}$ referring to the mass-ratio parameters of the up and down quarks, or the neutrinos and corresponding electrons (muons, tauons) etc. These mass parameters vary a lot from generation to generation, and from leptons to quarks. The Higgs field couples only  to the left handed gauge field of  $SU(2)_{L}$ and the hyper gauge field of $U(1)_{Y}$, i.e. $D^{\mu}\Phi^{\dagger}D_{\mu}\Phi$ with $D_{\mu}= (i\partial_{\mu} + Y g' A_{Y \mu} + gA_{L \mu})$ and $Y=\frac{1}{2}$ for the Higgs field. With a symmetry breaking potential $V=\frac{\lambda}{4}(\Phi^{\dagger}\Phi - v^{2})^{2}$ ($\lambda$ being positive), the Higgs field acquires a non-zero vacuum expectation value (VEV) $v$, resulting massive left-handed gauge bosons $W^{\pm}$ and $Z$, as well as massive fermions with masses equal to the mass ratio parameters multiplied by the VEV. The right handed gauge field does not exist. Actually, even the right handed global transformations of $SU(2)_{R}$ in the standard model are not symmetries of the Lagrangian, because the transformed Higgs field would not be a $SU(2)_{L}$ doublet any more. Besides the covariant kinetic terms of the fermions, the Yang-Mills Lagrangian also contains contribution from the field strength $-\frac{1}{4}\mathbf{tr}(F_{L\mu\nu}F_{L}^{\mu\nu} + F_{Y\mu\nu}F_{Y}^{\mu\nu}).$ No further gauge invariant terms are present. In the following it will be shown that this Higgs model can be derived from a pure gauge theory. 
\subsubsection{The gauge configuration for deriving the Standard-Model Higgs field} 
The starting point is the Lagrangian $L_{o}$ in Eq.(\ref{eq: Lo}) whose gauge group is  generated by the generators in $\mathbf{C}(6)^{C}$ specified in Eq. (\ref{eq: gen_nonabelian}). We postulate in some certain gauge, the chirality mixing gauge components are in the form:
\begin{equation} \label{eq: config_na}
\begin{split}
A_{mix,0} &=\phi(x) \sigma^{1}\otimes I_{2}\otimes I_{2}\\
A_{mix, 1}&=-i\phi(x) \sigma^{2}\otimes I_{2} \otimes \sigma^{1}\\
A_{mix, 2}&=i\phi(x) \sigma^{2}\otimes \sigma^{3} \otimes \sigma^{2}\\
A_{mix, 3}&=-i\phi(x) \sigma^{2}\otimes \sigma^{3} \otimes \sigma^{3}
\end{split}
\end{equation}

Under the Lorentz transformation $\Lambda_{\frac{1}{2}m}= \exp (\theta \sigma^{3}\otimes I_{2} \otimes \sigma^{**}) \cdot \exp(i\alpha I_{2}\otimes I_{2}\otimes \sigma^{1}) $ with $\sigma^{**}= \cos 2\alpha \sigma^{2} -\sin 2\alpha \sigma^{3}$ $A_{mix,\mu}$ are transformed into
\begin{equation}
\begin{split}
A_{mix,0} \Rightarrow & \cosh 2\theta \phi(x) \sigma^{1}\otimes I_{2} \otimes I_{2}\\
                 & +i\sinh 2\theta \phi(x) \sigma^{2} \otimes I_{2} \otimes \sigma^{**},\\
A_{mix,1} \Rightarrow & -i\phi(x) \sigma^{2}\otimes I_{2}\otimes \sigma^{1},\\
A_{mix,2} \Rightarrow & i\cosh2\theta \phi(x) \sigma^{2}\otimes \sigma^{3}\otimes \sigma^{**} \\
      &- \sinh 2\theta \phi(x) \sigma^{1}\otimes \sigma^{3} \otimes I_{2}, \\
A_{mix,3} \Rightarrow & -i \cos 2\alpha \phi(x) \sigma^{2}\otimes \sigma^{3}\otimes \sigma^{3} \\
&- i \sin 2\alpha \phi(x) \sigma^{2}\otimes \sigma^{3}\otimes \sigma^{2}.
\end{split}
\end{equation}
The visible chirality mixing gauge field components become
\begin{equation}
\begin{split}
A_{v,mix}= &(\cosh 2\theta-1) \phi(x)\sigma^{1}\otimes I_{2} \otimes I_{2} \\
 &+ (\cosh 2\theta-1) \cos 2\alpha  \phi(x)\sigma^{1} \otimes \sigma^{3} \otimes I_{2}
\end{split}
\end{equation}
Comparing this result to Eq.(\ref{eq: Higgs_SM}) the following mass ratio parameters are obtained:
\begin{equation}
\begin{split}
m_{a}=& 2\sinh^ {2}\theta\\
m_{b}=&2\sinh^{2}\theta \cos 2\alpha
\end{split}
\end{equation}
with $m_{a}+m_{b}=4\sinh^{2}\theta \cos^{2}\alpha$ and $m_{a}-m_{b}=4\sinh^{2}\theta \sin^{2}\alpha$ the mass ratios of fermions in each generation. Since $(\theta,\alpha)$ are hidden in the Gamma matrices $\Gamma_{m}^{\mu},$ they should viewed as a pair of quantum numbers. Let us calculate the Lorentz transformation explicitely
\begin{equation}
\begin{split}
\Lambda_{\frac{1}{2}m}=& \exp (\theta \sigma^{3}\otimes I_{2} \otimes \sigma^{*}) \cdot \exp(i\alpha I_{2}\otimes I_{2}\otimes \sigma^{1}) \\
                                     =&\cosh \theta \cos\alpha-\sinh\theta \sigma^{3}\otimes I_{2}\otimes \sigma^{*} \\
                                     & + i \cosh \theta \sin\alpha I_{2}\otimes I_{2}\otimes \sigma^{1}\\
                                     \Rightarrow & \exp(\beta \sigma)                               
\end{split}
\end{equation}
where $\sigma^{*}= \cos \alpha \sigma^{2} -\sin \alpha \sigma^{3},$ and the last line is added with the intention to find the generator $\sigma$ of the Lorentz transformation. But this leads to the discussion over three cases defined by the sign of the squared norm of $\sigma:$ $-\sinh^{2}\theta+ \cosh^{2} \theta \sin^{2}\alpha.$ The special case where the norm is zero puts constraint on $\theta$ and $\alpha.$ Unfortunately this does not fit well with the experimental data. On the other hand, it is curious to notice that the constraint $\cosh \theta \cos\alpha=\sinh\theta$ seems to be compatible with the known information about the masses of the leptons. With this assumption and the input of the masses of electron, muon, and tauon \cite{Wiki}, the masses of the neutrinos are estimated to be $\hat{m}_{\nu_{e}}=0.41ev,$ $\hat{m}_{\nu_{\mu}}=0.017Mev$ and $\hat{m}_{\nu_{\tau}}=4.92Mev.$ 

The rest of the gauge field components are postulated to take the following form: 

The left handed gauge field components:
\begin{equation}
A_{L\mu}=A_{L\mu}^{k}(x)\sigma^{L}\otimes \sigma^{k} \otimes I_{2} \  \ \textrm{with} \  \ k\in\{1,2, 3\}.
\end{equation}
with $A^{k}_{L\mu}(x)$ arbitrary real functions.
The right-handed gauge field components:
\begin{equation}
A_{R \mu}= a_{\mu}(x) \sigma^{R}\otimes \sigma^{3} \otimes I_{2} + C_{3R\mu}
\end{equation}
with $a_{\mu}(x)$ arbitrary functions of spacetime, and $C_{R \mu}=v_{\mu}  \sigma^{R}\otimes \sigma^{3}\otimes I_{2}$ with $v_{\mu}$ a constant vector. That is, the temporal component $A_{R \mu}$ is constrained to the third direction and shifted by a constant value $C_{3R\mu}.$ This constraint is only temporal and will be relaxed later.  The constant $v_{\mu}$ will serve to provide the VEV of the Higgs field. The gauge group generated by $\sigma^{R}\otimes \sigma^{3}\otimes I_{2}$ will be referred to as $U(1)_{R3}.$ The coupling constant, as is inherited from the $SU(2)_{R} \subset U(4) \subset G$ ($G$ is the group generated by $\mathbf{C(6)}^{C}$), must be $g$, the same as of $SU(2)_{L}.$  Transformations generated by $\sigma^{R}\otimes \sigma^{1}\otimes I_{2}$ and $\sigma^{R}\otimes \sigma^{2}\otimes I_{2}$ will be discussed later. 

\subsubsection{The interpretation of the $U(1)_{Y}$ hyper transformation}
Yet there is another $U(1)$ gauge sub-group generated by $I_{2}\otimes I_{2}\otimes I_{2},$ with the gauge field
\begin{equation}
A_{N \mu}=b_{\mu}(x) I_{2}\otimes I_{2} \otimes I_{2}.
\end{equation}
The gauge group generated by $I_{2}\otimes I_{2} \otimes I_{2}$ will be denoted by $U(1)_{N}.$ Since this generator belongs to the same Lie algebra and Clifford algebra with the other generators, it is natural to assume that the coupling constant is also $g.$

To stay aligned with the standard model, the $U(1)_{R3}$ and $U(1)_{N}$ transformations must go hand-in-hand:
\begin{equation}
\begin{split}
U(1)_{R3}=& \exp(i \alpha(x)\sigma^{R}\otimes \sigma^{3}\otimes I_{2})\\        
U(1)_{N}= & \exp( i n_{f} \alpha(x)I_{2}\otimes I_{2}\otimes I_{2}),
\end{split}
\end{equation}
where $n_{f}$ denotes the $U(1)_{N}$ charge carried by each generation of fermions. The lepton-doublets all have the same $U(1)_{N}$ charge $n_{l}=-1$ while quark pairs all having the same $U(1)_{N}$ charge $n_{q}=1/3.$ The right handed fermions in each generation all transform in the same way under $U(1)_{R3},$ as they must because this is derived from the $SU(2)_{R}$ transformation. The hyper charge of the Higgs field is due to its transformation property under $U(1)_{R3}$ which is explained in detail in appendix B. Since the chirality mixing generators commute with $U(1)_{N},$ the Higgs field does not carry $U(1)_{N}$ charges This arrangement also fixes the identification of the four Weyl fermions in $\Psi$ specified in Eq.(\ref{eq: Psi}) : $u$ refers to $u, c, t$ quarks and $\nu_{e}, \nu_{\mu}, \nu_{\tau}$ neutrinos; $d$ refers to $d, s, b$ quarks and $e, \mu, \tau$ the charged leptons. For clarity and illustration the assignments of the relevant charges are listed in the following table. It is clear that the hyper charge fits nicely as the sum of the $U(1)_{N}$ and $U(1)_{3R}$ charges.
\begin{center}
{\renewcommand{\arraystretch}{1.2}
\begin{tabular}{ |p{2cm}||p{2cm}|p{1.5cm}|p{1.5cm}|  }
 \hline
 \multicolumn{4}{c}{ The hyper charge assignment} \\
 \hline
    & $U(1)_{N}$ ($n_{f}$)& $U(1)_{R3}$ & $U(1)_{Y}$\\
 \hline
                     &~~ L ~~~  R   & L  \  ~   \ R & L \  ~   \ R\\
  \hline
  $\nu_{e}$, $\nu_{\mu}$, $\nu_{\tau}$   & ~~ -1\  ~ \  -1 & 0 \  ~   \  1& -1 \   ~ \ 0 \\

 e, $\mu$, $\tau$  &~~ -1\ ~  \  -1     & 0 \  ~  \ -1& -1 \  ~\  -2 \\
 u, c, t              &~~~$\frac{1}{3}$ \  ~ \ $\frac{1}{3}$ & 0 \  ~   \ 1& $\frac{1}{3}$ \ ~ \ $\frac{4}{3}$\\

 d, s, b            &~~~$\frac{1}{3}$ \ ~  \ $\frac{1}{3}$ & 0 \  ~  \ -1& $\frac{1}{3}$ \ ~  \ -$\frac{2}{3}$\\
 \hline
Higgs            &~~ \ ~ \ 0  & \ ~ \  1 & \ ~ \ 1 \\
 \hline
\end{tabular}}
\end{center}
Note that the values of hyper charge in the table are two times that given in the reference \cite{PS}, because the electric charge quantum number  in our system is defined as $Q=\frac{1}{2}(\sigma^{L}\otimes\sigma^{3}\otimes I_{2} + n_{f} I_{2}\otimes I_{2}\otimes I_{2}+ \sigma^{R}\otimes\sigma^{3}\otimes I_{2}),$ which is actually the same thing as in \cite{PS}.

Now there are two copies of $U(1)$ gauge fields associated to $U(1)_{N}$ and $U(1)_{R3}$ must be the same as the two transformations go hand in hand, i.e. $a_{\mu}(x)=b_{\mu}(x).$ But these gauge fields need to be scaled to $p_{\mu}(x)$ so that the two copies form a proper single $U(1)_{Y}$ gauge field. The scaling is done as follows:
\begin{equation}
\begin{split}
g a_{\mu}(x)\sigma^{R}\otimes \sigma^{3}\otimes I_{2}=& g' p_{\mu}(x)\sigma^{R}\otimes \sigma^{3}\otimes I_{2}\\
n_{f} g b_{\mu}(x)I_{2}\otimes I_{2}\otimes I_{2}=&n_{f} g' p_{\mu}(x) I_{2}\otimes I_{2}\otimes I_{2}\\
\end{split}
\end{equation}
which leads to the field strength part in the Lagrangian as: 
\begin{equation}
\begin{split}
&-\frac{1}{4} \mathbf{tr}[ (\partial_{\mu}a_{\nu}- \partial_{\nu}a_{\mu})\sigma^{R}\otimes \sigma^{3}\otimes I_{2}(\partial^{\mu}a^{\nu}- \partial^{\nu}a^{\mu})\sigma^{R}\otimes \sigma^{3}\otimes I_{2}\\
&+(\partial_{\mu}b_{\nu}- \partial_{\nu}b_{\mu})I_{2}\otimes I_{2}\otimes I_{2}(\partial^{\mu}b^{\nu}- \partial^{\nu}b^{\mu}) I_{2}\otimes I_{2}\otimes I_{2}]\\
=& -\frac{3}{8}(\partial_{\mu}a_{\nu}- \partial_{\nu}a_{\mu})(\partial^{\mu}a^{\nu}- \partial^{\nu}a^{\mu})\\
:=&-\frac{1}{4}\mathbf{tr} (\partial_{\mu}p_{\nu}- \partial_{\nu}p_{\mu})\tau\otimes I_{2}\otimes I_{2}(\partial^{\mu}p^{\nu}- \partial^{\nu}p^{\mu})\tau\otimes I_{2}\otimes I_{2} \\
=&-\frac{1}{8} (\partial_{\mu}p_{\nu}- \partial_{\nu}p_{\mu})(\partial^{\mu}p^{\nu}- \partial^{\nu}p^{\mu}),
\end{split}
\end{equation}
where $\mathbf{tr}\tau \otimes I_{2} \otimes I_{2}=1/2$ is assumed because the $U(1)_{Y}$ by our definition acts only on the left and right handed fermions doublets,  separately. The above calculation gives the scaling relation $p_{\mu} = \sqrt{3} a_{\mu},$ leading to $g^{'2}/g^{2}=1/3$ which agrees with the value of the Weinberg angle. Let's define $B_{\mu}=p_{\mu}\tau\otimes I_{2}\otimes I_{2}$ as the hyper gauge field for later convenience. 

With the new interpretation for the hyper charge, the electro-magnetic charge is then understood as the $U(1)_{N}$ charge $n_f$ plus the charge of $U(1)_{3}$ generated by $\sigma^{L}\otimes \sigma^{3}\otimes I_{2}+\sigma^{R}\otimes \sigma^{3}\otimes I_{2}= I_{2}\otimes \sigma^{3}\otimes I_{2}.$ It is curious to notice that this generator  $I_{2}\otimes \sigma^{3}\otimes I_{2}$ is exactly the bi-vector formed by the two non-spacetime generating basis of $\mathbf{C}(6)$ discussed before (this implies that the metric-signature of the extra two dimensions should be the same).

\subsubsection{The Lagrangian of the electroweak theory---derived from a pure gauge theory}
The current Lagrangian of the electro-weak sector in the standard model has gone through the most stringent tests. It serves as our best guide while searching for the correct theory. This leads us to the slight modification of the curvature part of the Yang-Mills Lagrangian:
\begin{equation}
-\frac{1}{8}\mathbf{tr}F_{\mu\nu}F^{\mu\nu} +h.c. \rightarrow -\frac{1}{8}\mathbf{tr}ZF_{\mu\nu}ZF^{\mu\nu} +h.c.
\end{equation}
where $Z=\sigma^{3}\otimes I_{2}\otimes I_{2}.$ The operator $Z$ commutes with the chirality keeping generators and anti-commutes with the chirality mixing generators, so that the sign of the kinetic term of the Higgs field can be made correct. Under a global or gauge transformation, $Z$ should be transformed as well, to keep everything invariant. 

Thus the gauge configuration postulated earlier gives the following Lagrangian of the electro-weak sector:
\begin{equation}
\begin{split}
L_{o} =& \sum_{m}\frac{1}{2}\Psi_{m}^{\dagger}h\Gamma_{m}^{\mu}(i\partial_{\mu} +A_{L\mu}+ B_{\mu} + A_{mix,\mu}) \Psi_{m}+h.c.\\
           & +\frac{1}{2}(D_{\mu}\Phi)^{\dagger}D^{\mu}\Phi - \frac{g^{2}}{3}[(\Phi^{\dagger}\Phi)^{2} - 2\tilde {v}^{2} \Phi^{\dagger}\Phi]\\
           &-\frac{1}{4} \mathbf{tr}( F_{L\mu\nu}F_{L}^{\mu\nu} + f_{Y\mu\nu}f_{Y}^{\mu\nu})
           \end{split}
\end{equation}
where $f_{Y\mu\nu}=\partial_{\mu}B_{\nu}-\partial_{\nu}B_{\mu}.$
Upon the restriction to $SU(2)_{L}\times U(1)_{Y},$ after performing the internal Lorentz transformation discussed earlier, this reduces to 
 \begin{equation}
\begin{split}
L_{t} =& \sum_{m}\Psi_{m}^{\dagger}h\Gamma^{\mu}(i\partial_{\mu} +gA_{L\mu}+ g'B_{\mu}) \Psi_{m}\\
           &+  \sum_{m}\Psi_{m}^{\dagger} g \Phi_{m}\Psi_{m} \\
            & +\frac{1}{2}(D_{\mu}\Phi)^{\dagger}D^{\mu}\Phi - \frac{g^{2}}{3}[(\Phi^{\dagger}\Phi)^{2} - 2\tilde {v}^{2} \Phi^{\dagger}\Phi]\\
           &-\frac{1}{4} \mathbf{tr}( F_{L\mu\nu}F_{L}^{\mu\nu} + f_{Y\mu\nu}f_{Y}^{\mu\nu} )  \\
           &+g^{2} v^{\mu} (A_{R\mu}^{3}-A_{L\mu}^{3}) \Phi^{\dagger}\Phi      
           \end{split}\end{equation}
where $\Phi_{m}$ is defined as
\begin{equation}
\Phi_{m}=\begin{pmatrix}
  0 &  \begin{split} m_{u}\bar{\Phi}^{\dagger}\\
       m_{d}\Phi^{\dagger}
  \end{split}\\
m_{u}\bar{\Phi}, m_{d}\Phi& 0
\end{pmatrix}
\end{equation}
with $\bar{\Phi}=i\sigma^{2}\Phi^{*}.$
The details of the calculation are presented in the appendix. An immediate observation is that there is an extra-term $g^{2}v^{\mu}(A^{3}_{L\mu}-A^{3}_{R\mu})\Phi^{\dagger}\Phi=gv^{\mu}(gA^{3}_{L\mu}-g'p_{\mu})\Phi^{\dagger}\Phi=g\sqrt{g^{2}+g^{'2}}v^{\mu}Z_{\mu}\Phi^{\dagger}\Phi$ with $v_{\mu}v^{\mu} = 4/3* (246 GeV)^{2}$ where $Z_{\mu}$ is the gauge field for the well-known $Z$ boson. We have not performed analysis on the quantum theory. Hopefully there are some processes that can help determine the values of $v_{\mu}.$ 

\subsubsection{The right-handed gauge bosons and the parity violation}
In the postulated gauge configuration, the gauge components $A^{1}_{R\mu}\sigma^{R}\otimes \sigma^{1}\otimes I_{2}$ and $A^{2}_{R\mu}\sigma^{R}\otimes \sigma^{2}\otimes I_{2}$ were suppressed. But there is no good reason that these two components should not be present. If they are included and the Lagrangian is recalculated, the following new term appears:
\begin{equation}
\begin{split}
&-g^{2}(v_{\mu}v^{\mu}A^{1}_{R \nu} A^{1\nu}_{R} + v_{\mu}v^{\mu}A^{2}_{R \nu} A^{2\nu}_{R} +\\
&-v_{\mu}v^{\nu}A^{1}_{R \nu} A^{1 \mu}_{R} - v_{\mu}v^{\nu}A^{2}_{R \nu} A^{2\mu}_{R} )
\end{split}
\end{equation}
where summation over $\mu$ and $\nu$ with $\mu \neq \nu$ is implied. 
When the gauge $A^{1,2}_{R0}=0$ is chosen, this resembles the mass terms of six massive bosons except that the sign appears to be wrong. Noticing that this term comes from the "electric field" of the field strength, when switching from the Lagrangian to the Hamiltonian the sign should not change. Thus they are indeed the mass terms of six massive bosons. The mass can be estimated to be at least 1.9 times that of the known $W^{\pm}$ gauge bosons.

This shows that they are too heavy to be discovered on the current colliders. Thus at the energy level in the previously conducted experiments they were not excited. This also explains why in the laboratory parity violation is observed. Reflections upon this analysis are inspiring for understanding the failure of observing a type of interaction in experiments. The interactions are described by a gauge theory. If the observation of a certain interaction is missing in experiment, it could just mean the corresponding gauge field (component) is difficult to get excited. The difficulty is most probably due to heavy energy-cost, which in a gauge theory is equivalent to large value of the curvature. In a non-abelian gauge theory the non-vanishing curvature can simply come from the mutual influence of several gauge components from the non-linear term, which often contributes significantly, in the field strength, leading to the effect that a constant gauge component in one direction prohibits excitation in the other directions. This is to a certain degree similar to the daily experience that a hanging curtain made of soft cloth cannot easily get wavy in the vertical direction while often made wavy by a breeze in the horizontal direction, because of the existence of the vertical gravitational acceleration.

\subsubsection{The topological term involving the Higgs field}
When the Higgs field is put in the gauge framework, it is possible to construct a topological term involving the Higgs field, namely the Pontryagin number or the Chern-Simons term:
\begin{equation} \label{eq: top}
\begin{split}
&\int \mathbf{tr} \epsilon^{\mu\nu\alpha\beta} F_{\mu\nu}F_{\alpha\beta}=\\
& 4\int \epsilon^{\mu\nu\alpha\beta}\mathbf{tr}\partial_{\mu}(A_{\nu}\partial_{\alpha}A_{\beta}-\frac{2i}{3}A_{\nu}A_{\alpha}A_{\beta})
\end{split}
\end{equation}
The topology of the group $U(8)_{C}$ is homeomorphic to $U(8)\times R^{64},$ so the gauge field is classified by $\pi_{3} U(8)\simeq Z.$ Since now the $U(1)_{Y}$ gauge field has been given a different interpretation, and the gauge field components associated to the chirality mixing, as well as that associated to the Lorentz group can be treated in the same gauge group, there arises difference in Eq.(\ref{eq: top}) compared to the intensively discussed calculation for the gauge group $SU(2)_{L}\times U(1)_{Y}$ in literature. With the gauge configuration postulated in Eq.(\ref{eq: config_na}) and thereafter, the Higgs field (the chirality mixing gauge components) makes no contribution to the topological term. When the gauge components associated to the Lorentz group are excited the Higgs field will make non-trivial contribution. This leads to modification to the gravitational anomaly. On the other hand, the $U(1)_{N}$ and the $U(1)_{3R}$ gauge components, as well as the background field do make a special contribution. Therefore, anomaly cancellation should be re-examined for discreetness, which might lead constraint on the fermion mass-ratios. 
\subsection{The abelian Higgs models}

The abelian Higgs model is defined by the following Lagrangian:
\begin{equation} \label{eq: aH}
\begin{split}
L_{A}=& \psi_{f}^{\dagger}\gamma^{0}\gamma^{\mu}(i\partial_{\mu}-e_{R}A_{R\mu}- e_{L}A_{L\mu})\psi_{f} + D^{\mu}\phi^{\dagger}D_{\mu}\phi \\
           &+m_{f} (\psi_{fR}^{\dagger} \phi \psi_{fL} + \psi_{fL}^{\dagger} \phi^{\dagger} \psi_{fR}) \\
           &- \frac{\lambda}{4!}(\phi^{\dagger}\phi- v^{2})^{2} -\frac{1}{4} \mathbf{tr}(F_{R \mu\nu}F_{R}^{\mu\nu}+F_{L\mu\nu}F_{L}^{\mu\nu})
\end{split}
\end{equation}
In this theory the Higgs field $\phi$ couples to both the left and right handed gauge fields of $U(1)_{R}$ and $U(1)_{L}$, i.e. $D_{\mu}=i\partial_{\mu} - e_{R}A_{R \mu} - e_{L}A_{L \mu}.$ In the Lagrangian only the charge units $e_{R}$ and $e_{L}$ are explicitly expressed. One can freely assign different charges to these fermions. There can be multiple Dirac fermions with different mass ratios $m_{f}$:
\begin{equation}
 \psi_{f}= \begin{pmatrix} \psi_{fR}\\
  \psi_{fL}\end{pmatrix}
  \end{equation}
Usually, no other gauge invariant terms are present in the theory. \\

\subsubsection{Deriving the abelian Higgs model from a gauge theory}
In this case the full gauge group is generated by the Lie algebra (also the Clifford algebra) $\mathbf{C}(1,3)$ that has been specified earlier. The gauge field is assumed to take the following form:\\
1) the chirality mixing gauge field components:
\begin{equation}
\begin{split}
A_{mix,0} &=\phi(x)  \sigma^{1}\otimes I_{2} \\
A_{mix, 1}&=i\phi(x)  \sigma^{2}\otimes \sigma^{1} \\
A_{mix, 2}&=i\phi(x)  \sigma^{2}\otimes \sigma^{2}  \\
A_{mix, 3}&=i\phi(x)  \sigma^{2}\otimes \sigma^{3}. 
\end{split}
\end{equation}
2) the chirality keeping gauge field components:
\begin{equation}
A_{keep, \mu} = a_{\mu}\sigma^{R}\otimes I_{2} + b_{\mu} \sigma^{L}\otimes I_{2} + v_{\mu}\sigma^{3}\otimes I_{2}
\end{equation}
That means a Left-Right symmetric theory will be obtained.  

Now just as in the non-abelian case when an internal Lorentz transformation is performed with $\Lambda_{\frac{1}{2}m}=\exp(\theta \sigma^{3} \otimes \sigma^{**}) \exp(i\alpha I_{2}\otimes \sigma^{1})$ with $\sigma^{**}= \cos 2\alpha \sigma^{2} -\sin 2\alpha \sigma^{3}$, these chirality mixing gauge components are transformed into:
\begin{equation}
\begin{split}
A_{mix,0} &\Rightarrow   \cosh 2\theta \phi(x) \sigma^{1}\otimes I_{2} + i\sinh 2\theta \phi(x) \sigma^{2}\otimes \sigma^{**}\\
A_{mix, 1}&\Rightarrow i\phi(x)  \sigma^{2}\otimes \sigma^{1} \\
A_{mix, 2}&\Rightarrow i\cosh 2\theta \phi(x)  \sigma^{2}\otimes \sigma^{**} - \sinh 2\theta \phi(x) \sigma^{1} \otimes I_{2}  \\
A_{mix, 3}&\Rightarrow i\cos 2\alpha \phi(x)  \sigma^{2}\otimes \sigma^{3}+  i\sin 2\alpha \sigma^{2}\otimes \sigma^{2}. 
\end{split}
\end{equation}
These field components would become visible as
\begin{equation}
A_{v,mix}=4\cosh^{2}\theta \cos^{2}\alpha \phi(x) \sigma^{1}\otimes I_{2}\\
\end{equation} 
which results in a mass ratio parameter $m =4\cosh^{2}\theta \cos^{2}\alpha.$  Alternative sign choices in $A_{mix,\mu}$ could give $m =4\sinh^{2}\theta \cos^{2}\alpha$ similar to that in the non-abelian case.

With this gauge configuration curvature term makes the following contribution to the Yang-Mills Lagrangian
\begin{equation}
\begin{split}
-\frac{1}{8} \mathbf{tr} ZF_{\mu\nu}ZF^{\mu\nu}+h.c.=& -\frac{1}{4} f_{L\mu\nu}f_{L}^{\mu\nu} -\frac{1}{4} f_{R\mu\nu}f_{R}^{\mu\nu} \\
      &+\frac{1}{2}(D_{\mu}\varphi)^{\dagger} D^{\mu}\varphi \\
      &- \frac{e^{2}}{3}(\varphi^{\dagger}\varphi)^{2} +\frac{e^{2}}{2} v_{\mu}v^{\mu} \varphi^{\dagger}\varphi\\
     & +e^{2}v^{\mu} (b_{\mu}-a_{\mu})\varphi^{\dagger}\varphi
      \end{split}
\end{equation}
with $D_{\mu}\phi=(i\partial_{\mu} - e_{R}a_{\mu} - e_{L}b_{\mu})\phi$ with $e_{R}$ and $e_{L}$ arbitrary numbers for each generation of fermions. Combined with the fermionic part this gives the Lagrangian $L_{o}.$ Following the same procedure as in the non-abelian case, $L_{t}$ can be derived, exactly in the form as expected. 

\subsubsection{The varying mass and the topology}
In the same way as in the non-abelian case, it can be argued that the mass of fermions may develop a dependence on the space-time location when the background field $v_{\mu}\sigma^{3}\otimes I_{2}$ varies in the spacetime. As a non-abelian model may describe the phenomena in some low-temperature superconducting system. It is hoped that this can provide a testing ground for this hypothesis. 

If the geometric algebra (Lie algebra) is not complexified, the full symmetry group generated by $\mathbf{C}(1,3)$ is $U(1,3)$ which has its largest compact subgroup $U(1)\times SU(2)_{H}\times SU(2)_{S}.$ The subgroup $SU(2)_{H}$ is generated by $\sigma^{k}\otimes I_{2}$ acting on the handedness of the fermions, while $SU(2)_{S}$ is generated by the Lorentz rotations $I_{2}\otimes \sigma^{k}$ acting on the spin of fermions.  As a manifold, there exist automorphisms on $U(1,3)$ that transform the sub-manifolds $SU(2)_{H}$ and $SU(2)_{S}$ to each other. However, taking the physical constraint into account, we are limited to only gauge transformations, none of which can transform $SU(2)_{H}$ to $SU(2)_{S}$ as they commute with each other. Thus the topology of the gauge field, considered as maps from the boundary of the spacetime ($S^{3}$) to the group $U(1,3)$, should be classified by $Z\times Z.$ This implies that the abelian Higgs model has a richer topological content than the non-abelian model. Suppose the algebra $\mathbf{C}(1,3)$ is indeed complexified, but there is an energy barrier to excite the extra gauge components, the above argument remains safe. But at higher energies, the topology of the gauge field would become $\pi_{3} U(4)\simeq Z.$
 \section{Problems and further research directions}
 Although the approach presented previously exhibits a lot of attractive features, it also brought up a number of confusing problems that invite serious reflections. Let's make some shallow remarks as a first attempt to address these problems.
 
1)  In both the abelian and non-abelian cases the symmetry group of the Lagrangians are enlarged considerably and become non-compact, especially in the non-abelian case the Lie algebra is duplicated by a necessary complexification. There are three problems to be addressed regarding the two symmetry groups.

First, the non-compactness of a symmetry generator is accompanied by the "negative-energy" problem. In the curvature term $\frac{1}{8}\mathbf{tr}ZF_{\mu\nu}ZF^{\dagger \mu\nu} +h.c.$ the Killing form is not positive definite for all symmetry generators. In particular, it is negative for the non-compact generators. This would lead to negative energy when switching from the Lagrangian formalism to the Hamiltonian formalism. A usual trick to circumvent this problem is to choose a positive definite metric by hand. In both the abelian and non-abelian cases, we adopted the indefinite metric to calculate the Lagrangian as the indefinite metric made the signs of the terms in the Higgs sector in accord with the known Higgs models. However, as a gauge theory, when switching to the Hamiltonian formalism, the Hamiltonian obtained would not be the same as if the Higgs field were treated as a unique, non-gauge, dynamical variable. Our speculation is, the correct curvature term in the Lagrangian is 
\begin{equation}
-\frac{1}{4}\mathbf{tr} \mathcal{I}F_{\mu\nu}\mathcal{I}^{-1} F^{\dagger \mu\nu},
\end{equation} 
where $\mathcal{I}$ is the identity matrix with the transformation rule as a spinor metric. This solves the negative energy problem. Then it is noticed that the resulted Hamiltonian does not produce the $SU(2)_{L}$ symmetry-breaking Higgs potential necessary to generate masses for particles. A possible interpretation is that the theory should have been treated as constrained system. The gauge configurations postulated to derive the Higgs models are subjected to constraints. It is expected that the symmetry breaking potential emerges if the constraints are properly treated. 

Thus the correct Higgs potential obtained by adopting $\frac{1}{8}\mathbf{tr}F_{\mu\nu}F^{\dagger \mu\nu} +h.c.$ is an accidental result which happens to agree with the standard model. Whether or not there is a justification is not known yet. But it is certain that this comes with the condition that the chirality mixing gauge component identified as the Higgs should be treated as a non-gauge related independent field. This is clearly the feature of a constrained system. 

Second, there is a mismatch between the two cases: in the non-abelian case, the Lie algebra is almost complexified by the principle of closeness, as result leading to complexified geometric algebra of 6 dimensions. In particular the Lorentz group is complexified. However, in the abelian case the Lorentz group is not complexified. This mismatch is confusing as the spacetime in both cases is the same one and they must possess the same symmetry. Possibly the geometric algebra $C(1,3)$ in the abelian case is also complexified to $C(1,3)^{C}.$  We notice that in the non-abelian case, it is the introduction of two-copies of the Dirac field in each generation that forces us to see the interweaving relationship between the matter and the spacetime and the necessity to complexify the Lie algebra and the geometric algebra. In contrast, in the abelian case the matter field content is too simple to reveal this necessity. 

At last, the "over-size" of the symmetry groups in both cases raises serious concerns, as in reality only a very small fraction of the interactions described by the two full symmetry groups are observed. If some of symmetries were not to be gauged, what should be the guiding principle to discern which to be gauged and which not? One direction to look at is the effect of the transformations on the spinor metric $h.$ Under some symmetry transformations the spinor metric does not stay invariant. If the invariance of $h$ is set as a criterion for whether or not to gauge the symmetry, unfortunately transformations generated by $\sigma^{L}\otimes \sigma^{k}\otimes I_{2}$ would be excluded from the gauged symmetries. This cannot be the case. In fact, $\sigma^{L}\otimes \sigma^{k}\otimes I_{2}$ commutes with the products $h\Gamma^{\mu}$ which guarantees the products as whole entities invariant. But the products $h\Gamma^{\mu}$ would be altered by a Lorentz transformation. These arguments suggest that a simple criterion related to $h$ cannot be working. To find the true criterion, a clear understanding of the nature of the spinor metric is probably the key. Another direction is to assume that all the symmetries can be gauged, but excited with various energy-costs. After all, all the symmetries can be equally interpreted as coordinate-transformations in the internal vector space (spinors jointly formed by spin, handedness and isospin) with a norm defined. A transformation also induces a new norm. Thus making the coordinate transformation local and introduce the corresponding connection field component is a very natural thing. With the background $C_{3R}$ being postulated, excitations along some directions are going to be more energy costing. Further, it may be the case that the gauge field component associated to a non-compact transformation is more energy costing than that of a compact one. These two factors would substantially limit the number of observable interactions at the usual energy level.\\

 2).The second confusing aspect concerns the chirality mixing sector of the gauge configurations postulated for deriving the Higgs models. Clearly, this configuration represents a constrained excitation in the chirality mixing sector. On the other hand the chirality mixing transformations (both global and local) are forbidden. Otherwise, the structure of the Higgs field would be destroyed. Regarding the constraint it is reasonable to suspect that there is some force (energy condition) that constrains the chirality mixing gauge components. It is appealing to ask about the nature of the constraining force as well as the mechanism how it works. Inspired by the work in \cite{Witten} and \cite{MF}, it is tempting to seek a spacetime symmetry imposed on the gauge field. But no success has been achieved in this line except noticing that in the abelian case, the chirality mixing gauge components in the chosen gauge is a grade-1 vector symmetrically aligned with respect to the four directions, in the sense of geometric algebra. In the non-abelian case, it is impossible to simultaneously identify the four generators $\sigma^{1}\otimes I_{2}\otimes I_{2}$ and $i\sigma^{2}\otimes \sigma^{3}\otimes I_{2}$ as four vectors in the geometric algebra $\mathbf{C}(1,6)^{C}$. Instead, they can be identified as one vector and three tri-vectors. Regarding the symmetry transformation, there is no obvious reason to exclude the chirality mixing gauge transformations since the (constrained) gauge field is already there. A possible explanation to this is that those symmetry transformations are not forbidden. But they would make it more difficult to interpret what we observe. That is to say, the current theory has shaped our mind to understand the physical world in terms of mass and the known types of interactions. The chirality mixing symmetry transformations would make the concept of mass hard to comprehend, thus we are more comfortable to stay in the gauge where there is a simple concept of mass that aids to understand physical phenomena in a simple way.   \\
 
 3) There is also a curious fact about the $U(1)_{Y}$ gauge transformations in the non-abelian case. As was noted, these symmetry transformations cannot be generated by a single generator, instead they are formed by the simultaneous transformations of $U(1)_{N}$ and $U(1)_{R3},$ as the leptons and quarks are assigned different values of $U(1)_{N}$ charges $n_{f}$. This fact means that the gauge field components associated to both groups are excited. Since the $U(1)_{N}$ and $U(1)_{R3}$ in principle can be gauged independently, there arises a question then, why must these two $U(1)$ transformations act in a hand-in-hand way? What happens to the other combination of the two gauge field components? It is observed that, had the quarks presented in triples and the anti-particles are observed, the baryons and the leptons would have the same $U(1)_{N}$ charges, then the $U(1)_{Y}$ symmetry could be generated by a single generator, $I_{2}\otimes I_{2}\otimes I_{2} + \sigma^{R}\otimes \sigma^{3}\otimes I_{2},$ and the question about the excitation of the other combination of the two gauge field components would become less pressing. In the labs indeed the quarks are observed in triples. This suggests that perhaps that the $U(1)_{Y}$ is indeed generated by a single generator, and it is some mysterious properties of the quarks that force us to look at it in a complicated way. 
    
 4)Speculation on the concept of varying mass dependent on spacetime locations.
Analysis into the mechanism for generating the fermion masses shows that the key for the non-vanishing fermion mass values is the constant background gauge components $C_{3R\mu}$ postulated in the gauge configuration. If the background gauge component were non-constant, or even zero, a lot of our observations would be altered. In particular, the masses of particles would develop a dependence on the spacetime locations. This is equivalent to speculate that the VEV of the Higgs field in the standard model is spacetime dependent. Then we wonder: why must there be a non-vanishing constant background? Now that the Higgs field is identified as the chirality mixing gauge field components that need to be treated together with the Lorentz group, it is not far-reaching to speculate that the presence of the postulated background may be due to some gravitational condition. Further, it seems bizarre to have a constant non-zero background gauge connection with zero curvature everywhere in the universe. Since $C_{3R\mu}$ is related to the electro-magnetic symmetry, this hints towards a non-zero electric-magnetic field in some distant region where the gauge component is non-constant. Presence of electric-charges must be accompanied by the presence of matters. Altogether, this reasoning leads us to the speculation that there might be some unknown matter distribution in some remote area of the universe. An alternative possibility is to understand the non-vanishing background as a topological effect, like kinks and instantons, constrained not to vanish by a non-trivial topological condition.\\

\section{Appendices}
Throughout, we adopt the particle-physicists convention for the spacetime metric and the Weyl representation of the gamma matrices, which are:
\begin{gather}
g_{\mu \nu}=\begin{bmatrix}  1 & 0 & 0 & 0 \\
0 & -1 & 0 & 0 \\
0 & 0 & -1 & 0 \\
0 & 0 & 0 &  -1
\end{bmatrix}, 
\end{gather} 
for $\mu, \nu= 0,1,2,3$ and 
\begin{equation}
\begin{split}
\gamma^{0}=\begin{bmatrix}  0 & 0 & 1 & 0 \\
0 & 0 & 0 & 1 \\
1 & 0 & 0 & 0 \\
0 & 1 & 0 & 0
\end{bmatrix}, & \gamma^{1}=\begin{bmatrix}  0 & 0 & 0 & 1 \\
0 & 0 & 1 & 0 \\
0 & -1& 0 & 0 \\
-1 & 0 & 0 & 0
\end{bmatrix},\\
\gamma^{2}=\begin{bmatrix}  0 & 0 & 0 & -i \\
0 & 0 & i& 0 \\
0 & i & 0 & 0 \\
-i & 0 & 0 & 0
\end{bmatrix},
& \gamma^{3}=\begin{bmatrix}  0 & 0 & 1 & 0 \\
0 & 0 & 0 & -1 \\
-1 & 0 & 0 & 0 \\
0 & 1 & 0 & 0
\end{bmatrix}
\end{split}
\end{equation}
leading to the expression of the $\Gamma^{\mu} s$ given in Eq.(\ref{eq: Gamma}). 
The Pauli matrices are:

\begin{gather}
\sigma^{1}=\begin{pmatrix} 0 & 1 \\
1 &0 \end{pmatrix} , \   \  \sigma^{2}=\begin{pmatrix} 0 & -i \\
i &0 \end{pmatrix} , \    \ \sigma^{3}=\begin{pmatrix} 1 & 0 \\
0 & -1 \end{pmatrix} \end{gather} 

\subsection{Appendix A --- The Higgs field as the $SU(2)_{L} \times U(1)_{Y}$ representation}

The Higgs field in the Yukawa coupling terms in the standard model is identified as the visible chirality mixing gauge components $A_{v,mix}=m_{a}\Phi_{a}+ m_{b}\Phi_{b}$ with $m_{a}, m_{b}$ and $\Phi_{a}, \Phi_{b}$ defined in Eq.(\ref{eq: Config_Higgs})
This can be put in the form $A_{v,mix}= m_{u}\phi_{s} X^{s} \otimes \sigma^{0}+ m_{d}\phi_{s}Y^{s}\otimes \sigma^{0}$ with $s\in \{ 1, 2, 3, 4\}$ and
\begin{equation}
\begin{split}
X^{1}= \sigma^{2}\otimes \sigma^{2}-\sigma^{1}\otimes \sigma^{1} ,& X^{2}= \sigma^{2}\otimes \sigma^{1}+\sigma^{1}\otimes \sigma^{2}, \\
X^{3}= \sigma^{1}\otimes I_{2}+\sigma^{1}\otimes \sigma^{3}, & X^{4}= \sigma^{2}\otimes \sigma^{3}+ \sigma^{2}\otimes I_{2}, \\
Y^{1}= \sigma^{2}\otimes \sigma^{2}+ \sigma^{1}\otimes \sigma^{1}, & Y^{2}= \sigma^{2}\otimes \sigma^{1}- \sigma^{1}\otimes \sigma^{2}, \\
Y^{3}= \sigma^{1}\otimes I_{2}- \sigma^{1}\otimes \sigma^{3}, & Y^{4}=\sigma^{2}\otimes \sigma^{3}- \sigma^{2}\otimes I_{2}. 
\end{split}
\end{equation}
It is worth noting that the $X^{s}$ and $Y^{s}$ are all orthogonal to each other with respect to the norm $\mathbf{tr},$ and each of them has the norm $\sqrt{2}/2.$ Further, when $\sigma^{0}$ in the definition $A_{mix}= m_{u}\phi_{s} X^{s} \otimes \sigma^{0}+ m_{d}\phi_{s}Y^{s}\otimes \sigma^{0}$ is replaced by $\sigma^{k},$ the previous statement as well as the following discussion still holds. This will be useful in the next appendix for computing the Higgs sector of the Lagrangian.
 
Now the following maps can be defined: 
\begin{equation} \label{eq: maps}
\begin{split}
\phi_{s}X^{s}& \rightarrow \Phi=\begin{pmatrix} 
\phi_{1}+ i\phi_{2}\\
\phi_{3} - i\phi_{4}
\end{pmatrix},\\
\phi_{s}Y^{s} &\rightarrow \bar{\Phi}: =i\sigma^{2} \Phi^{*}=\begin{pmatrix} 
\phi_{3}+ i\phi_{4}\\
-\phi_{1} + i\phi_{2}
\end{pmatrix}\end{split}
\end{equation} 
These maps correspond to the construction of a spinor from the Rotors in the Clifford algebra. By computing the commutators $[\sigma^{L}\otimes \sigma^{k}, X^{s}]$, $[\sigma^{L}\otimes \sigma^{k}, Y^{s}]$, $[\sigma^{R}\otimes \sigma^{3}, X^{s}]$ and $[\sigma^{R}\otimes \sigma^{3}, Y^{s}]$ it is straightforward to verify the following:
\begin{equation} \label{eq: trans_Phi}
\begin{split}
&\exp(i\alpha \sigma^{R}\otimes \sigma^{3}) \phi_{s}X^{s}\exp(-i\alpha \sigma^{R}\otimes\sigma^{3}) \rightarrow e^{i\alpha} \Phi \\
&\exp(i\alpha \sigma^{L}\otimes \sigma^{k}) \phi_{s}X^{s}\exp(-i\alpha \sigma^{L}\otimes \sigma^{k}) \rightarrow e^{i\alpha\sigma^{k}} \Phi \\
\end{split}
\end{equation}
and 
\begin{equation} \label{eq: trans_barPhi}
\begin{split}
&\exp(i\alpha \sigma^{R}\otimes \sigma^{3}) \phi_{s}Y^{s}\exp(-i\alpha \sigma^{R}\otimes\sigma^{3}) \rightarrow e^{-i\alpha} \bar{\Phi} \\
&\exp(i\alpha \sigma^{L}\otimes \sigma^{k}) \phi_{s}Y^{s}\exp(-i\alpha \sigma^{L}\otimes \sigma^{k}) \rightarrow e^{i\alpha\sigma^{k}} \bar{\Phi} \\
\end{split}
\end{equation}
Thus we see that $\{ \sigma^{R}\otimes \sigma^{3}, \sigma^{L}\otimes \sigma^{k} \}$ form the Lie-algebra basis of the $U(2)$ group of $\phi$, and $\{-\sigma^{R}\otimes \sigma^{3}, \sigma^{L}\otimes \sigma^{k} \}$ that of $\Phi. $ Note that the transformation generated by $\sigma^{R}\otimes I_{2}$ rotate $X^{s}$ and $Y^{s}$ in the same direction. That's why $\phi$ and $\Phi$ acquire opposite phases (there is an extra complex conjugation operation on $\Phi.$) This responds to the difference in the electric charges of the two Dirac fermions in a pair such as electron and electric neutrino.

Based on the discussion above, $A_{v,mix}$ can be rewritten as:
\begin{equation}
\Phi=\begin{pmatrix}
  0 &  \begin{split} m_{u}\bar{\Phi}^{\dagger}\\
       m_{d}\Phi^{\dagger}
  \end{split}\\
m_{u}\bar{\Phi}, m_{d}\Phi& 0
\end{pmatrix}
\end{equation}
It is clear that under the rotations generated by the $\sigma^{R}\otimes \sigma^{3}$ and $\sigma^{L}\otimes \sigma^{k},$  $\Phi$ transforms just as $\Phi$ and $\bar{\Phi}$ do under the corresponding $U(2)$ transformations. Further, the map defined in Eq. {\ref{eq: maps}} and the results in Eq.s (\ref{eq: trans_Phi}) and (\ref{eq: trans_barPhi}) lead to the mapping rule of the covariant derivative of $\phi_{s}X^{s}$ and $\phi_{s}Y^{s}$ as:
\begin{equation}
\begin{split}
&\partial_{\mu}(\phi_{s}X^{s}) - ig[A^{k}_{\mu}\sigma^{L}\otimes \sigma^{k}+ A^{0}_{\mu}\sigma^{R}\otimes\sigma^{3}, \phi_{s}X^{s}] \\
&\rightarrow \partial_{\Phi}-ig (A^{k}_{\mu}\sigma^{k}+A^{0}_{\mu} I_{2})\Phi :=D_{\mu}\Phi\\
&\partial_{\mu}(\phi_{s}Y^{s}) - ig[A^{k}_{\mu}\sigma^{L}\otimes \sigma^{k}+ A^{0}_{\mu}\sigma^{R}\otimes\sigma^{3}, \phi_{s}Y^{s}]\\
& \rightarrow \partial_{\Phi}-ig (A^{k}_{\mu}\sigma^{k}+A^{0}_{\mu} I_{2})\Phi =D_{\mu}\Phi\\\end{split}
\end{equation}

It is interesting to look at the effect of the transformations generated by the other chirality keeping generators on $\phi_{s}X^{s}$ and $\phi_{s}Y^{s}.$ Calculation of $[\sigma^{R, L}\otimes I_{2}, X^{s}]$, $[\sigma^{R,L}\otimes I_{2}, Y^{s}]$, $[\sigma^{R}\otimes \sigma^{1,2}, X^{s}]$ and $[\sigma^{R}\otimes \sigma^{1,2}, Y^{s}]$ reveals the following
\begin{equation}
\begin{split}
&\exp(i\alpha \sigma^{R}\otimes I_{2}) \phi_{s}X^{s}\exp(-i\alpha \sigma^{R}\otimes I_{2}) \rightarrow e^{i\alpha} \Phi \\
&\exp(i\alpha \sigma^{R}\otimes I_{2}) \phi_{s}Y^{s}\exp(-i\alpha \sigma^{R}\otimes\sigma^{3}) \rightarrow e^{i\alpha} \bar{\Phi} \\
&\exp(i\alpha \sigma^{L}\otimes I_{2}) \phi_{s}X^{s}\exp(-i\alpha \sigma^{R}\otimes I_{2}) \rightarrow e^{-i\alpha} \Phi \\
&\exp(i\alpha \sigma^{L}\otimes  I_{2}) \phi_{s}Y^{s}\exp(-i\alpha \sigma^{L}\otimes \sigma^{1}) \rightarrow e^{-i\alpha} \bar{\Phi} \\
\end{split}
\end{equation}
and transformations generated by $\sigma^{R}\otimes \sigma^{1,2}$ will mix $X^{s}$ and $Y^{s}.$ The transformation law shows that $\sigma^{R,L} \otimes I_{2}$ (more precisely $\sigma^{3}\otimes I_{2}$) rotate $X^{s}$ and $Y^{s}$ in the opposite directions, thus $\Phi$ and $\bar{\Phi}$ acquires the same phase under the same rotation. This would violate the "complex-conjugation" relation between $\bar{\Phi}$ and $\Phi,$ this is the rotation generated by $\sigma^{3}\otimes I_{2}$ does not preserve the relation $\bar{\Phi} = i\sigma^{2} \Phi^{*}.$ On the other hand, $A_{mix}$ the parameters $m_{a}$ and $m_{b}$ as coefficients of $\Phi_{a}$ and $\Phi_{b}$ should stay the same after the symmetry transformation, in order to define a consistent doublet $\Phi.$ These arguments exclude the $\sigma^{R,L}\otimes I_{2}$ and $\sigma^{R}\otimes \sigma^{1,2}$ as valid symmetry generators of $\Phi.$ Of course, the logic here is that $m_{a}$ and $m_{b}$ are determined first and then consistency in the definition of $\Phi$ is demanded. Had the logic been reversed, a right-handed Higgs doublet $\Phi'$ could be defined as well.


\subsection{Appendix B--- Calculation of the Higgs sector of the Lagrangian}

In this appendix the Higgs sector in the Lagrangian is derived for both the abelian and non-abelian Higgs models, from a pure gauge theory with constraints on the gauge configuration postulated. 

1) The non-abelian Higgs model

The curvature contribution to the Yang-Mills Lagrangian is hypothesised to be $-\frac{1}{8}\mathbf{tr} ZF_{\mu\nu}ZF^{\mu\nu}+h.c.$ with $Z=\sigma^{3}\otimes I_{2}\otimes I_{2}.$ When the gauge field is grouped into chirality keeping and mixing components as: $A_{\mu}= A_{L\mu}+A_{R\mu}+C_{3R\mu}+A_{N\mu}+\eta_{\mu}$ with $C_{3R\mu} =v_{\mu} \sigma^{R}\otimes\sigma^{3}\otimes I_{2}$, where $v_{\mu}$ is a constant vector, and 
\begin{equation}
\begin{split}
&A_{L\mu}=A^{k}_{L\mu}(x) \sigma^{L}\otimes \sigma^{k}\otimes I_{2}   \    \  k \in \{1,2,3 \}\\
&A_{R\mu}=a_{\mu}(x)\sigma^{R}\otimes\sigma^{3}\otimes I_{2}\\
&A_{N\mu}=a_{\mu}(x)I_{2}\otimes I_{2}\otimes I_{2}\\
&\eta_{0}=\phi(x)  \sigma^{1}\otimes I_{2}\otimes I_{2}\\
&\eta_{1}=-i\phi(x)  \sigma^{2}\otimes I_{2} \otimes \sigma^{1} \\
&\eta_{2}=i\phi(x)  \sigma^{2}\otimes \sigma^{3}\otimes \sigma^{2} \\
&\eta_{3}=-i\phi(x)  \sigma^{2}\otimes \sigma^{3}\otimes \sigma^{3} 
\end{split}
\end{equation}

Then the field strength can be written as follows:
\begin{equation}
\begin{split}
F_{\mu\nu} =& \partial_{\mu} A_{L\nu} - \partial_{\nu}A_{L\mu}   + \partial_{\mu} A_{R\nu} - \partial_{\nu}A_{R\mu}\\
  &+\partial_{\mu} A_{N\nu} - \partial_{\nu}A_{N\mu}+ \partial_{\mu} \eta_{\nu} - \partial_{\nu}\eta_{\mu}-ig[A_{\mu},A_{\nu}] \\
=&\partial_{\mu} A_{L\nu} - \partial_{\nu}A_{L\mu} -i g[A_{L\mu},A_{L\nu}] + \partial_{\mu} A_{R\nu} - \partial_{\nu}A_{R\mu}\\
&+ \partial_{\mu} A_{N\nu} - \partial_{\nu}A_{N\mu}+ (\partial_{\mu}\eta_{\nu} - ig[A_{L\mu}+A_{R\mu}, \eta_{\nu}])\\
& -(\partial_{\nu}\eta_{\mu} - ig[A_{L\nu}+A_{R\nu}, \eta_{\mu}])\\
& -ig[C_{3R\mu}, \eta_{\nu}]-ig[\eta_{\mu}, C_{3R\nu}]-ig[\eta_{\mu}, \eta_{\nu}]\\
:=& F_{L\mu\nu}+ f_{R\mu\nu} +f_{N\mu\nu} +D_{\mu}\eta_{\nu} - D_{\nu}\eta_{\mu}\\
&-ig([C_{3R\mu}, \eta_{\nu}]-[C_{3R\nu}, \eta_{\mu}])-  i g[\eta_{\mu}, \eta_{\nu}]\\
\end{split}
\end{equation}
where gauge field strength are defined for the subgroups, $SU(2)_{L}, U(1)_{R}$ and $U(1)_{N}$, and the covariant derivative of $\eta_{\mu}$ as $D_{\mu}\eta_{\nu}=\partial_{\mu}\eta_{\nu} - ig[A_{L\mu}+A_{R\mu}, \eta_{\nu}].$ 
Since $Z$ commutes with all the chirality keeping generators and anti-commute with the chirality mixing generators, the curvature conjugated by $Z$ is:
\begin{equation}
\begin{split}
F_{\mu\nu} = &F_{L\mu\nu}+ f_{R\mu\nu} +f_{N\mu\nu} -D_{\mu}\eta_{\nu} + D_{\nu}\eta_{\mu}\\
&+ig([C_{3R\mu}, \eta_{\nu}]-[C_{3R\nu}, \eta_{\mu}])-  i g[\eta_{\mu}, \eta_{\nu}]
\end{split}
\end{equation}
Then we have
\begin{equation}
\begin{split}
-\frac{1}{8}&\mathbf{tr}ZF_{\mu\nu}ZF^{\mu\nu} +h.c.=-\frac{1}{4}\{\mathbf{tr}F_{L\mu\nu}F_{L}^{\mu\nu} +\mathbf{tr}f_{R\mu\nu}f_{R}^{\mu\nu}\\
&+\mathbf{tr} f_{N\mu\nu}f_{N}^{\mu\nu} -\mathbf{tr}(D_{\mu}\eta_{\nu}-D_{\nu}\eta_{\mu}) (D^{\mu}\eta^{\nu}-D^{\nu}\eta^{\mu}) \\
&-g^{2}\mathbf{tr}[\eta_{\mu}, \eta_{\nu}][\eta^{\mu}, \eta^{\nu}]\\
&+g^{2}\mathbf{tr}([C_{3R\mu}, \eta_{\nu}]-[C_{3R\nu}, \eta_{\mu}])([C_{3R}^{\mu}, \eta^{\nu}]-[C_{3R}^{\nu}, \eta^{\mu}])\\
&-2ig\mathbf{tr}(F_{L\mu\nu}+f_{R\mu\nu}+f_{N\mu\nu}) [\eta^{\mu}, \eta^{\nu}]\\
&+2ig\mathbf{tr}(D_{\mu}\eta_{\nu} - D_{\nu}\eta_{\mu})([C_{3R}^{\mu}, \eta^{\nu}]-[C_{3R}^{\nu}, \eta^{\mu}]) \}\\
&=-\frac{1}{4}\mathbf{tr}F_{L\mu\nu}F_{L}^{\mu\nu} -\frac{1}{4}\mathbf{tr}f_{R\mu\nu}f_{R}^{\mu\nu}-\frac{1}{4}\mathbf{tr} f_{N\mu\nu}f_{N}^{\mu\nu}\\ &+\frac{3}{2}(D_{\mu}\Phi)^{\dagger}D^{\mu}\Phi - 3g^{2}(\Phi^{\dagger}\Phi)^{2} +\frac{3}{2}g^{2} v_{\mu}v^{\mu} \Phi^{\dagger}\Phi\\
& +3g^{2} v^{\mu} (A_{R\mu}^{3}-A_{L\mu}^{3}) \Phi^{\dagger}\Phi
\end{split}
\end{equation}
in deriving which the normalisation $\mathbf{tr} I_{8} =\frac{1}{8} \textrm{tr} I_{8}=1$ has been applied. The field $\Phi$ is to be identified as the $SU(2)_{L}$ doublet in the standard model and in the currently chosen gauge $\Phi= \begin{pmatrix} 0\\\phi(x)  \end{pmatrix}.$ Its covariant derivative is defined in the same way as in the previous appendix. To match precisely with the parameters in the standard model, the field $\phi$ needs to be rescaled and the constant vector $v_{\mu}$ must be set as follows: 
\begin{equation}
\begin{split}
&\phi(x) \rightarrow \sqrt{\frac{1}{3}}\phi(x)\\
&v_{\mu}v^{\mu} = \frac{4}{3} \tilde{v}^{2} =  \frac{4}{3}* (246 GeV)^{2}
\end{split}
\end{equation}
Calculating the relevant terms in $-\frac{1}{4}ZF_{\mu\nu}ZF^{\mu\nu}$ gives the Higgs sector of the Lagrangian
\begin{equation}
\begin{split}
L_{\Phi} =& \frac{1}{2}(D_{\mu}\Phi)^{\dagger}D^{\mu}\Phi - \frac{g^{2}}{3}[(\Phi^{\dagger}\Phi)^{2} - 2\tilde {v}^{2} \Phi^{\dagger}\Phi] \\
&+g^{2} v^{\mu} (A_{R\mu}^{3}-A_{L\mu}^{3}) \Phi^{\dagger}\Phi
\end{split}
\end{equation}
If the right-handed gauge field components $A_{R\mu}^{1,2}\sigma^{R}\sigma^{1,2}\otimes I_{2}$ are restored in the theory, the term
$\frac{g^{2}}{4}\mathbf{tr}([C_{3R\mu}, A_{R\nu}]-[C_{3R\nu}, A_{R\mu}])([C_{3R}^{\mu}, A_{R}^{\nu}]-[C_{3R}^{\nu}, A_{R}^{\mu}])$ would produce two terms $-g^{2}v_{\mu}v^{\mu}A^{1}_{R\nu}A_{R}^{1\nu}+g^{2}v_{\mu}v^{\nu}A^{1}_{R \nu} A^{1 \mu}_{R}$ and $-g^{2}v_{\mu}v^{\mu}A^{2}_{R\nu}A_{R}^{2\nu}+g^{2}v_{\mu}v^{\nu}A^{2}_{R \nu} A^{2 \mu}_{R} $ where in the summation $\mu \neq \nu.$ When the gauge $A_{R0}^{1,2}=0$ is chosen and in the frame where $v_{k}=0$ this gives  $+g^{2}v_{0}v^{0}A^{1}_{k}A^{1}_{k}$ and $+g^{2}v_{0}v^{0}A^{2}_{k}A^{2}_{k}$. Given that the vector $v_{\mu}$ must be timelike, the sign of these two terms seems to be opposite to that of a bosonic mass term in a Lagrangian. However, these terms are essentially the $E_{k}E^{k}$ part of electric-magnetic Lagrangian. Thus in the Hamiltonian, they would not change the sign and forming mass terms with correct signs. Thus, these two terms together with $\frac{1}{2}(D_{\mu}\Phi)^{\dagger}D^{\mu}\Phi$ produce six mass terms with the least mass value, 
\begin{equation}
(\frac{4}{3}+\frac{1}{2})g^{2} \tilde{v}^{2} A_{Rk}^{a}A_{Rk}^{a} \textrm{ for } a\in \{1,2\} \textrm{ and } k\in\{1, 2, 3\}
\end{equation}
While the ${W^{\pm}}$ field components would acquire a mass coefficient $\frac{1}{2}g^{2}\tilde{v}^{2}$ from $\frac{1}{2}(D_{\mu}\Phi)^{\dagger}D^{\mu}\Phi$, it can be predicted that these right-handed weak gauge bosons have masses at least $1.9$ times of the left-handed $W^{\pm}$ bosons, too heavy to be discovered at the currently running colliders. Also they are too heavy to be excited which results in the disparity between the left-handed and right-handed weak interactions. \\

2)The abelian Higgs model

To reconstruct the abelian Higgs model, let us assume that the gauge field takes up the following configuration:
$A_{\mu}= A_{L\mu}+A_{R\mu}+C_{3R\mu}+\eta_{\mu}$ with $C_{3R\mu} =v_{\mu} \sigma^{R}\otimes I_{2}$ (actually $C_{L\mu} =v_{\mu} \sigma^{L}\otimes I_{2}$ and  $C_{\mu} =v_{\mu} \sigma^{3}\otimes I_{2}$ equally do the job) and 
\begin{equation}
\begin{split}
&A_{L\mu}=a_{\mu}(x) \sigma^{L}\otimes I_{2}\\
&A_{R\mu}=b_{\mu}(x)\sigma^{R}\otimes I_{2}\\
&\eta_{0}=\frac{1}{\sqrt{3}}\phi(x)  \sigma^{1}\otimes I_{2}\\
&\eta_{k}=\frac{i}{\sqrt{3}}\phi(x)  \sigma^{2}\otimes \sigma^{k}. \\
\end{split}
\end{equation}
Precisely in the same way, this gauge configuration leads to the following:
\begin{equation}
\begin{split}
-\frac{1}{8}\mathbf{tr}ZF_{\mu\nu}ZF^{\mu\nu}+h.c.=&-\frac{1}{4}\mathbf{tr}f_{L\mu\nu}f_{L}^{\mu\nu} -\frac{1}{4}\mathbf{tr}f_{R\mu\nu}f_{R}^{\mu\nu}\\
&+\frac{1}{2}(D_{\mu}\varphi)^{\dagger}D^{\mu}\varphi \\
&- \frac{e^{2}}{3}(\varphi^{\dagger}\varphi)^{2} +\frac{e^{2}}{2} v_{\mu}v^{\mu} \varphi^{\dagger}\varphi\\
& +g^{2} v^{\mu}(b_{\mu}-a_{\mu})\varphi^{\dagger}\varphi
\end{split}
\end{equation}
where $Z$ now equals to $\sigma^{3}\otimes I_{2}$ and the normalisation condition $\mathbf{tr} I_{4} = \frac{1}{4}\textrm{tr} I_{4}=1$ has been used. 
The new variable $\varphi = \varphi_{1}(x) +i \varphi_{2}$ in this gauge takes the value $\phi(x)$ and $D_{\mu}\varphi = \partial_{\mu}\varphi -ie(a_{\mu}-b_{\mu})\varphi$ and it is subjected to both the left and right symmetry transformations.\\

\begin{acknowledgments}
The author enjoyed intimate discussions with Dr. Liu Yachao during the period of this work. This work was supported by The Special Foundation for theoretical physics (Grant No. 11847168), and partly supported by The Natural Science Youth Foundation of Shaanxi (Grant No. 2019JQ-739).  
\end{acknowledgments}


\end{document}